\documentclass[onecolumn,authoryear]{els-mrw} 

\usepackage{amsmath,amssymb,amsfonts,amsthm,makeidx,graphicx}
\usepackage{txfonts}
\usepackage{helvet}
\usepackage{xcolor}
\usepackage{hyperref}

\def\lum {erg\,s$^{-1}$}
\def\flux {erg\,s$^{-1}$cm$^{-2}$}
\def\ss {s\,s$^{-1}$}
\def\cm2{cm$^{-2}$}

\begin{document}

\chapter{The zoo of isolated neutron stars}\label{chapINS}

\author[1,2]{Alice Borghese}
\author[3,4,5]{Francesco Coti Zelati}

\address[1]{\orgname{INAF--Osservatorio Astronomico di Roma}, \orgaddress{via Frascati 33, I-00078 Monteporzio Catone, Italy}}
\address[2]{\orgname{European Space Agency},\orgdiv{European Space Astronomy Centre},\orgaddress{Camino Bajo del Castillo s/n, 28692 Villanueva de la Ca\~nada, Madrid, Spain}}
\address[3]{\orgname{Institute of Space Sciences (ICE, CSIC)}, \orgdiv{Campus UAB}, \orgaddress{Carrer de Can Magrans s/n, E-08193 Barcelona, Spain}}
\address[4]{\orgname{Institut d'Estudis Espacials de Catalunya (IEEC)}, \orgaddress{08860 Castelldefels (Barcelona), Spain}}
\address[5]{\orgname{INAF--Osservatorio Astronomico di Brera}, \orgaddress{Via Bianchi 46, I-23807 Merate (LC), Italy}}

\articletag{Chapter Article tagline: update of previous edition,, reprint..}

\maketitle


\begin{abstract}[Abstract]
Discovered over fifty years ago, neutron stars exhibit a remarkable variety of behaviors depending on their age, magnetic field strength, rotational dynamics, emission mechanisms, and surrounding environments. This diversity in their observational manifestations has led astronomers to classify neutron stars into numerous categories, much like wandering through a zoo and admiring various species. 
This chapter focuses on isolated (i.e., non-accreting) neutron stars. We review extensive observational results of rotation-powered pulsars, magnetars, X-ray dim isolated neutron stars and central compact objects, highlighting the unique properties and recent discoveries of these neutron star classes. We briefly touch on theoretical models that have significantly advanced our understanding of these classes, emphasizing the valuable insights they provide into the underlying physics of neutron stars.

\end{abstract}

Keyword -- Compact objects (288);
Magnetars (992);
Magnetic fields (994);
Millisecond pulsars (1062);
Neutron stars (1108);
Optical pulsars(1173);
Pulsars (1306);
Radio pulsars (1353);
Rotation powered pulsars (1408).


\begin{glossary}[Nomenclature]
\begin{tabular}{@{}lp{34pc}@{}}

$P$ (Spin period) & The time it takes for a neutron star to complete one rotation. \\

$\dot{P}$ (Spin period derivative) & The rate at which the spin period of a neutron star increases over time. \\

$\tau_{\rm c}$ (Characteristic age) & An estimate of the age of a neutron star, derived from its spin period and spin period derivative. \\

$B_{\rm dip}$ (Dipolar magnetic field) & The strength of the neutron star magnetic field at the poles, derived from its spin period and \\
& spin period derivative. \\

PSR (Pulsar) & A magnetized, rotating neutron star that emits beams of electromagnetic radiation. \\

RPP (Rotation-Powered Pulsar) & A pulsar whose emission is powered by the loss of rotational energy due to braking caused \\

& by their magnetic fields. \\

MSP (Millisecond Pulsar) & A sub-class of rotation-powered pulsars that have been spun up to millisecond periods due to \\

& the accretion of matter from a companion star. \\

RRAT (Rotating Radio Transient) & A neutron star that emits sporadic and brief bursts of radio waves, unlike regular pulsations \\

& of typical radio pulsars.\\

Magnetar (Magnetic Star) & A neutron star powered by the decay and instabilities of an ultra-strong magnetic field. \\

XDINS (X-ray Dim Isolated Neutron Star) & A thermally emitting neutron star detected primarily in soft X-rays with nearly thermal spectra,\\

& characterized by their faint optical and ultraviolet counterparts and lack of detected radio emissions. \\

CCO (Central Compact Object) & A weakly magnetized neutron star located at the center of a young supernova remnant, emitting \\

& soft X-rays with a thermal spectrum and showing no counterparts at other wavelengths. \\

\end{tabular}
\end{glossary}

\section*{Key points}
\begin{itemize}
    \item Introduction to the historical context and discovery of neutron stars.
    \item Overview of different types of isolated neutron stars, focusing on their observational properties.
    \item Explanation of the $P$-$\dot{P}$ diagram and its significance in classifying neutron stars.
    \item Description of rotation-powered pulsars, including their emission mechanisms and observational characteristics.
    \item Discussion on magnetars, their unique features, and emission processes.
    \item Examination of X-ray dim isolated neutron stars and their properties.
    \item Insights into central compact objects and their peculiar properties.
\end{itemize}

\section{Introduction}
\label{ins:intro}

The existence of neutron stars was first proposed by Baade and Zwicky in 1934 \citep{Baade1934} as the final stage in the life of a massive star following a supernova explosion: 

\begin{center}
   {\it "We have tentatively suggested that the super-nova process represents the transition of an ordinary star into a neutron star."} 
\end{center}

It took about 30 years to detect the first neutron star. Jocelyn Bell Burnell, then a Ph.D. student, discovered radio pulses from an unknown source in the sky using data from the Mullard Radio Astronomy Observatory. These pulses lasted $\sim$0.3\,s each and repeated every 1.337\,s\footnote{Actually, the first neutron star observed was Scorpius X-1 \citep{Giacconi1962}, a powerful X-ray source outside the Solar system. However, its classification as a neutron star powered by mass accretion is challenging due to the absence of detectable X-ray pulsations.}. Due to their extreme regularity, these radio signals were initially interpreted as radial pulsations from a white dwarf or a neutron star \citep{Hewish68}. Shortly after, \cite{pacini68} and \cite{gold68} independently suggested that these signals could be emitted by a spinning neutron star with a strong magnetic field, where particles are accelerated to relativistic velocities and emit beams of electromagnetic radiation. By then, around 20 objects with similar properties had been observed, including a source in the Crab nebula and one in the Vela supernova remnant. It became clear that a new class of celestial objects had been discovered, and the term {\it pulsar} was coined. The smoking gun linking pulsars to rotating neutron stars was the measurement of a spin-down in the period of the Crab pulsar. As noted by Pacini and Gold, a rotating object would slow down, whereas a vibrating one would not. Moreover, the location of some pulsars within supernova remnants further supported the initial suggestions made by Baade and Zwicky. Since then, the number of known pulsars has increased significantly. Nowadays, more than 3500 pulsars have been discovered, either isolated or in binary systems \citep[see the ATNF Pulsar Catalogue \url{https://www.atnf.csiro.au/research/pulsar/psrcat/};][]{Manchester05}. This number continues to grow thanks to more sensitive radio surveys, all-sky X-/gamma-rays monitors, and multi-wavelength observations targeting supernova remnant candidates. In fact, radio pulsars represent just one possible manifestation of the neutron star population: neutron stars have been found to emit pulsations across the entire electromagnetic spectrum, from radio to gamma-rays, and some are even radio-quiet.  

\begin{figure}[t]
\centering
\includegraphics[width=0.75\textwidth]{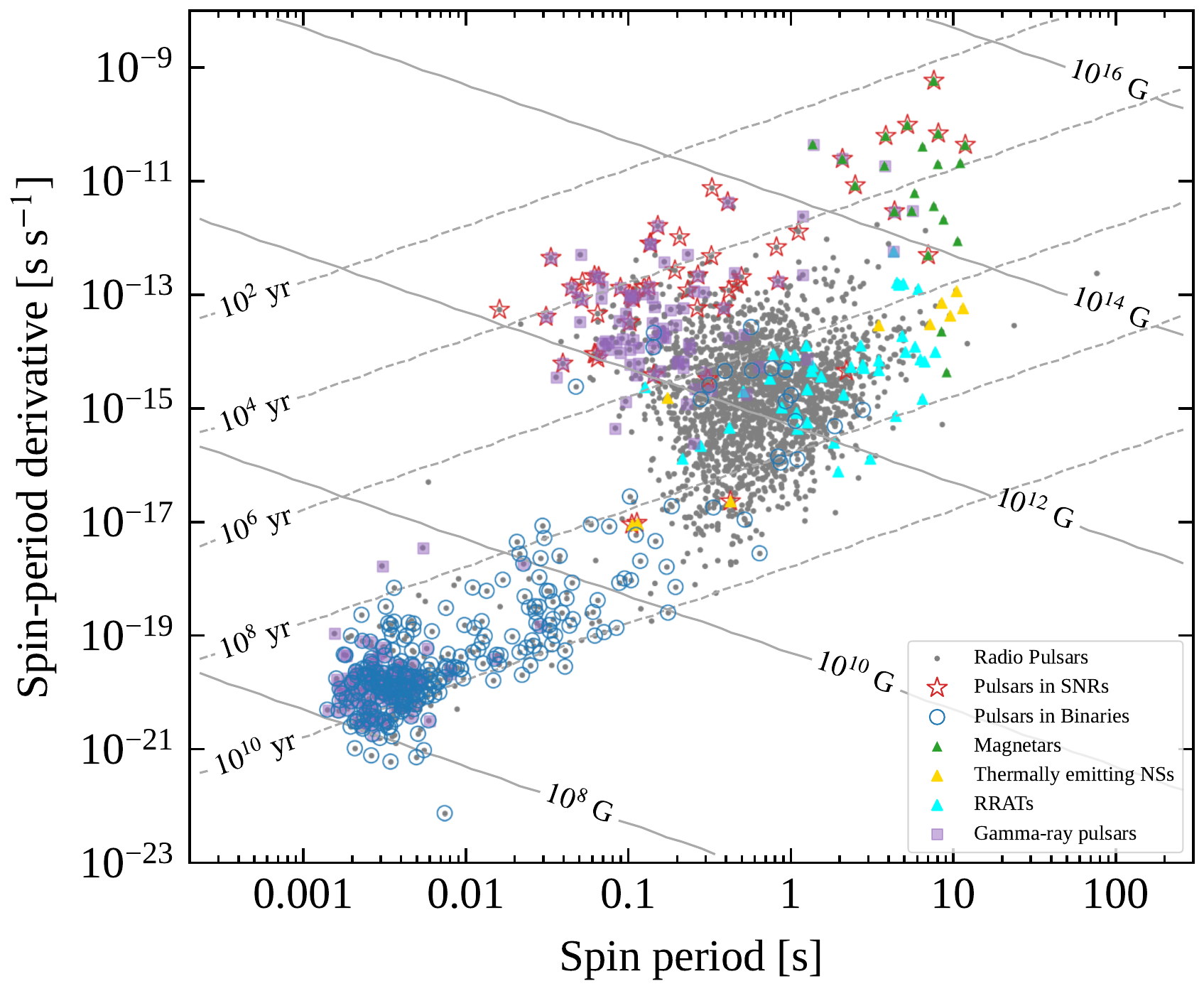}
\caption{Diagram showing the spin period derivative $\dot{P}$ as a function of the spin period $P$ for detected neutron stars (data from the ATNF Pulsar Catalogue). Grey dots and light blue triangles represent rotation-powered pulsars (RPPs) and rotating radio transients (RRATs), respectively. Green triangles and yellow triangles indicate magnetars and thermally emitting neutron stars, i.e. X-ray dim isolated neutron stars (XDINSs) and central compact objects (CCOs). Also plotted are lines of constant dipolar magnetic field (solid lines) and characteristic age (dashed lines), as derived from equations\,\ref{eq:B} and \ref{eq:tau}. Image credit: M. Ronchi.}
\label{fig:ppdotdiagram}
\end{figure}

Neutron stars exhibit diverse observational properties based on the primary source of energy that supplies their emission.
For instance, rotation-powered pulsars (RPPs) derive their energy from the loss of rotational kinetic energy due to braking caused by their magnetic fields. In contrast, accretion-powered neutron stars emit energy from the release of gravitational binding energy as material transferred from a companion star impacts their surface -- indeed these objects are always found in binary systems. Internal heat represents another potential energy reservoir, which could be residual heat from star formation or could result from surface reheating by external sources. Magnetically powered neutron stars typically have X-ray luminosity exceeding their rotational power and show no signs of accretion. Their emission is driven by the decay and instabilities of their strong magnetic fields. These emission mechanisms categorize neutron stars into different classes, although these mechanisms are not mutually exclusive. In this chapter, we will focus on {\it isolated} neutron stars, defined as neutron stars that do not accrete matter from a companion star
\citep[for a review see e.g.,][and references therein]{borghese23}.
In the following sections, we will detail the observational properties of RPPs (Sec.\,\ref{ins:psr}), magnetars (Sec.\,\ref{ins:magnetar}), X-ray dim isolated neutron stars (XDINSs, Sec.\,\ref{ins:xdin}) and central compact objects (CCOs, Sec.\,\ref{ins:cco}). The $P-\dot{P}$ diagram, shown in Figure\,\ref{fig:ppdotdiagram}, where $P$ is the spin period and $\dot{P}$ is its first derivative, is the simplest way to visualize the different groups of neutron stars. From measurements of $P$ and $\dot{P}$, we can derive estimates of key physical parameters for an isolated neutron star. According to the magnetic dipole braking model \citep[see e.g.,][]{pacini67}, a neutron star can be considered a rotating magnetic dipole in vacuum. Assuming the emission process is dipole radiation and that the neutron star spin-down is dominated by magnetic field torque, with the neutron star being an orthogonal rotator (i.e., rotational and magnetic axes are orthogonal), the strength of the dipolar magnetic field at the pole can be expressed as:
\begin{equation}
\label{eq:B}
    B_{\rm dip} \approx 6.4 \times 10^{19} (P \dot{P})^{1/2}\,{\rm G}.
\end{equation}
This formula assumes standard parameters for a neutron star: a mass $M_{\rm NS}$=1.4$M_\odot$ and a radius $R_{\rm NS}$=10\,km.
Under these same assumptions and further assuming that the current $P$ is significantly larger than its initial value at birth, the so-called characteristic age of the neutron star can be derived as:
\begin{equation}
\label{eq:tau}
    \tau_{\rm c} = \frac{P}{2\dot{P}}.
\end{equation}
It is worthy to note that $\tau_{\rm c}$ does not necessarily provide a reliable estimate of the true age of the neutron star because of the many assumptions and simplifications.
However, for most pulsars, it remains the only estimate available.




\section{Rotation-powered pulsars} \label{ins:psr}

Rotation-powered pulsars (RPPs) are detected through their pulsed emissions across multiple electromagnetic bands, from radio to very-high energy gamma-rays. These pulsars are powered by the rotation of their magnetic fields and populate a significant region of the $P$-$\dot{P}$ diagram (Figure\,\ref{fig:ppdotdiagram}). Over 3500 of these pulsars are known (see the ATNF Pulsar Catalogue; \citealt{Manchester05}). While most pulsars cease emitting as they cross the so-called ``death line''\footnote{This is a theoretical boundary in the $P$-$\dot{P}$ diagram beyond which the spin-down power of a pulsar is no longer able to sustain the processes producing detectable radio emission.} at some point during their evolutionary path, a sub-class of quickly rotating RPPs, dubbed millisecond pulsars (MSPs), are often still detectable beyond typical pulsar lifespans \citep{msp2022}. Currently, around 500 MSPs, defined as pulsars with spin periods shorter than 10\,ms, are known. The widely accepted scenario to explain their formation involves the accretion of matter from a companion star, which spins up the pulsar to millisecond periods \citep{Alpar1982}.

Radio pulsars are the most numerous type of RPPs, detected primarily through their periodic radio emission. This emission is generated by the acceleration of charged particles along the magnetic-field lines, producing beams that sweep across the Earth as the pulsar rotates. The precise timing of these pulses provides valuable insight into neutron star physics, the properties of dense matter within their interiors, the interstellar medium, and general relativity \citep{Wex2020}. 
A peculiar sub-class of radio pulsars are Rotating RAdio Transients (RRATs), which emit sporadic and brief bursts of radio waves, unlike the regular pulsations of typical radio pulsars. Discovered in 2006, RRATs have periods that range from hundreds of milliseconds to several seconds. Their sporadic nature makes them challenging to detect, and their underlying emission mechanism is not yet fully understood \citep{Keane2011}. Another intriguing discovery is the existence of nulling and intermittent radio pulsars, which exhibit sudden and temporary cessations in their emission lasting from a few seconds to several hours or even days. This phenomenon, first observed in the 1970s, has been detected in about 10\% of known pulsars, but its cause remains unclear. Nulling and intermittent pulsars provide an important laboratory for understanding the pulsar emission mechanism and the conditions under which it can switch on and off. 
The phenomenon of ``giant pulses'', observed in a small subset of radio pulsars, represents yet another peculiar aspect of pulsar emission. These giant pulses are extremely short, intense bursts of radio waves, which can be thousands of times stronger than the regular pulses from the same pulsar. First identified in the Crab pulsar, giant pulses have since been detected in a few other pulsars, suggesting a connection to the underlying mechanisms of pulsar magnetospheres and emission processes \citep{Kuzmin2007}. 
Recent discoveries have significantly expanded our understanding of the diversity among radio pulsars, while also challenging current theories of radio emission mechanisms and neutron star evolution. Notable examples include the 23.5-s radio pulsar PSR\,J0250$+$5854 \citep{Tan2018} and the 75.9-s radio pulsar PSR\,J0901$-$4046 \citep{Caleb2022}, which exhibits unique spectro-temporal properties. Additionally, two extremely peculiar periodic radio sources have been identified: GLEAM-X\,J162759.5$-$523504.3 \citep{HurleyWalker2022} and GPM\,J1839$–$10 \citep{HurleyWalker2023}. The former is a transient source that pulses every 18.18\,min and features polarized emission spanning a broad frequency range, with pulse widths varying between 30 and 60\,s and evolving on hourly timescales; the latter is a long-period radio transient that pulses every 21\,min and has been active for at least three decades. Its pulses vary in brightness, last between 30 and 300\,s, and exhibit quasi-periodic substructure. The observed properties of these sources, along with their location in the $P$-$\dot{P}$ diagram beyond the conventional pulsar death line, imply that our understanding of the pulsar death line and the conditions under which pulsars can emit radio waves is far from complete. Hence, theoretical models require significant revisions to accommodate these phenomena. Additionally, these discoveries indicate a larger and more diverse population of radio pulsars than previously known, suggesting that many similar sources may remain undetected due to observational biases.

Rotation-powered X-ray pulsars exhibit a wide range of X-ray luminosities and spectral characteristics, with typical X-ray efficiencies ranging from 10$^{-5}$ to 10$^{-3}$ of their spin-down power. This means that only a very small fraction of the energy lost as the pulsar slows down is converted into X-ray radiation. The X-ray emission from these pulsars can be broadly categorized into thermal and non-thermal components. The thermal component typically originates from the surface of the neutron star, particularly from hot spots at the magnetic poles heated by returning particles from the magnetosphere. This thermal emission is generally observed in the form of soft X-rays. On the other hand, the non-thermal component arises from the magnetosphere of the pulsar, where particles are accelerated to relativistic speeds, emitting high-energy radiation through synchrotron and curvature radiation processes. This non-thermal emission spans a broad spectrum, including hard X-rays and gamma-rays. Indeed, many RPPs are significant emitters of gamma rays.  Since its launch in 2008, the \emph{Fermi Gamma-ray Space Telescope} has been instrumental in significantly increasing the number of known gamma-ray pulsars and advancing our understanding of gamma-ray pulsars. As of now, it has identified approximately 340 gamma-ray pulsars and candidates, with 294 of these detected at GeV energies \citep{Smith2023}. These pulsars are often young and energetic, and can emit thousands of times more energy in gamma-rays than in radio waves. The gamma-ray spectra of pulsars generally show a cut-off around a few GeV, which indicates the maximum energy to which particles can be accelerated in their magnetospheres. Pulse profiles of gamma-ray pulsars commonly exhibit two peaks per rotation, corresponding to the emission regions near the magnetic poles where particles are accelerated and emit high-energy radiation. Very high energy (VHE) emission, reaching into the TeV range, has been detected by telescopes like H.E.S.S., MAGIC, and VERITAS from a few pulsars. Notable examples include the Crab Pulsar, detected up to 1.5\,TeV, and the Vela Pulsar, observed at over 100\,GeV. VHE emission is thought to arise from mechanisms like synchrotron radiation and inverse Compton scattering, where relativistic electrons boost photon fields to TeV energies \citep{Harding2021}. These objects are thus crucial for understanding the mechanisms of high-energy particle acceleration and emission in the magnetospheres of neutron stars.


Finally, rotation-powered optical pulsars are less studied compared to their radio, X-ray, and gamma-ray siblings \citep{Shearer2018}. As a matter of fact, following the early detection of optical emission from the Crab pulsar in 1969, only a few such pulsars have been identified to date, including Vela, Geminga, PSR\,B0540$-$69, and PSR\,B0656$+$14. The main reason for the limited number of known optical (and infrared) pulsars is their intrinsic faintness. In fact, the luminosity of these pulsars in the optical waveband is inversely proportional to the tenth power of the pulsar period, as predicted by \cite{pacini71}. Early observations of the Crab pulsar showed significant optical polarization, supporting the theory that the optical emission results from incoherent synchrotron radiation. Additionally, correlations between optical and radio emissions have been detected during giant radio pulses (where optical emission was enhanced by $\approx$3\%), suggesting the presence of an overlap in the emission zones for optical and radio photons.

\section{Magnetars} 
\label{ins:magnetar}

Historically recognized as two separate classes, anomalous X-ray pulsars and soft gamma-ray repeaters are nowadays believed to be different observational manifestations of the same type of object: a neutron star endowed with an ultra-strong magnetic field whose decay and instabilities power the emission, hence the name {\it magnetar} \cite[a portmanteau for magnetic star, see e.g.,][]{Duncan92}. The distinction between the two groups reflects how these sources were discovered: soft gamma-ray repeaters were identified through short, repeated bursts in the hard X-ray/soft gamma-ray band \cite[e.g.,][]{kouveliotou93}, while anomalous X-ray pulsars were discovered due to their high persistent X-ray luminosity \cite[$L_{\rm X}\approx10^{34}-10^{36}$\,\lum; e.g.,][]{vanparadijs95}. As of May 2024, about 30 magnetars have been discovered\footnote{See the McGill Magnetar Catalog \url{https://www.physics.mcgill.ca/~pulsar/magnetar/main.html} \citep{Olausen14}.}. All reside in the Galactic plane at low latitudes, except for two sources found in the Magellanic Clouds. Magnetars are slow rotators ($P\sim1-12$\,s), and they spin down over timescales of a few thousand years ($\dot{P}\sim10^{-13}-10^{-11}$\,\ss). These timing parameters imply $\tau_{\rm c}\sim10^3-10^5$\,yr and $B_{\rm dip}\sim10^{14}-10^{15}$\,G, making magnetars the strongest magnets in the Universe. Their persistent X-ray luminosity, $L_{\rm X}\approx10^{32}-10^{36}$\,\lum, is generally higher than their rotational energy loss rate \cite[see e.g.,][]{Cotizelati18}. 

\subsection*{Persistent emission}
Magnetars emit over a broad range of energies, from soft to hard X-rays, reaching $\sim$150--200\,keV \cite[see e.g.,][]{Gotz06}. The persistent emission in the soft X-ray range (0.5--10\,keV) is adequately described by a thermal component, typically modeled as a blackbody with a temperature of $kT\sim0.3-1$\,keV. In some cases, an additional component is required, either a hotter blackbody ($kT\sim1-2$\,keV) or a power law with a photon index of $\Gamma\sim2-4$. This phenomenological model is interpreted within a framework where the thermal emission originates from the hot surface of the neutron star and may be  affected by magnetospheric effects, such as resonant cyclotron scattering. Soft seed photons emitted from the surface are boosted to higher energies through repeated scatterings with charged particles flowing in a twisted magnetosphere, resulting in the formation of a power-law tail. The hard X-ray ($>10$\,keV) component is effectively represented by a power-law spectrum that is flatter compared to the soft X-ray range, characterized by a photon index of $\Gamma\sim0.5-2$ \citep{enoto17}. Upper limits in the hundreds of keV and MeV bands suggest that the hard X-ray spectra must break or bend between $\sim$300\,keV and 1\,MeV \cite[see e.g.,][]{Gotz06}. The process responsible for these hard tails is still poorly understood, but the prime candidate is resonant Compton up-scattering of soft thermal photons by a population of highly relativistic electrons threaded in the magnetosphere (see \citealt{Wadiasingh18} and references therein). 
Over the past few years, X-ray polarimetric observations of magnetars have provided significant insights into the physical conditions and behaviors of these ultra-magnetized neutron stars \citep{Taverna2024}. These observations have revealed that X-ray radiation from magnetars is highly polarized, with a degree of polarization ranging from approximately 15-20\% at lower energies (2--4\,keV) to more than 80\% at higher energies (8\,keV). This high degree of polarization indicates the presence of strong magnetic fields, and is consistent with the resonant Compton scattering model: the mode-dependent cross sections of the resonant Compton scattering process result in a higher fraction of X-mode photons, leading to significant polarization. The differing polarization patterns among magnetars indicate that the thermal emission mechanisms may vary. For instance, the emission could originate directly from the condensed surface of the star or be reprocessed in a thin, magnetized atmospheric layer above the crust. Interestingly, a peculiar 90-degree swing in the polarization angle has been detected in the magnetar 4U\,0142+61. This phenomenon can be explained by the propagation of photons in two normal modes --ordinary and extraordinary -- again suggesting the presence of ultra-strong magnetic fields and supporting the twisted magnetosphere model. Although no definitive evidence of vacuum birefringence was found, the phase-dependent behavior of the polarization angle strongly suggested its presence.

Exploring magnetar emissions in the optical and infrared regimes is challenging due to their intrinsic faintness at these wavelengths and their locations in heavily absorbed regions of the Galactic plane. Despite these difficulties, counterparts have been identified for approximately one-third of known magnetars \citep[see][and references therein]{Chrimes2022}. The associations are unambiguous for three sources, since optical pulsations at the X-ray period have been detected \cite[see e.g.,][]{Dhillon09}. In the context of the magnetar scenario, the optical and infrared emissions are believed to arise from non-thermal processes within the inner magnetosphere. Specifically, these emissions are likely generated by curvature radiation emitted by relativistic electrons moving along closed magnetic field lines \citep{beloborodov07}. However, a comprehensive model explaining these phenomena has yet to be developed.
Recent observations of the magnetar 4U 0142+61 indicate that its mid-infrared spectrum is better described by an absorbed power-law model rather than an absorbed blackbody model, pointing indeed to a nonthermal origin for the infrared emission. A comparison with simultaneous X-ray data show that the infrared spectrum is significantly softer than the X-ray spectrum, implying different sites for the emissions in the two bands in the magnetosphere \citep{Hare2024}.

\subsection*{Transient activity}
\begin{figure}[t]
\centering
\includegraphics[width=0.95\textwidth, trim={0.2cm 8cm 0.2cm 0cm},clip=true]{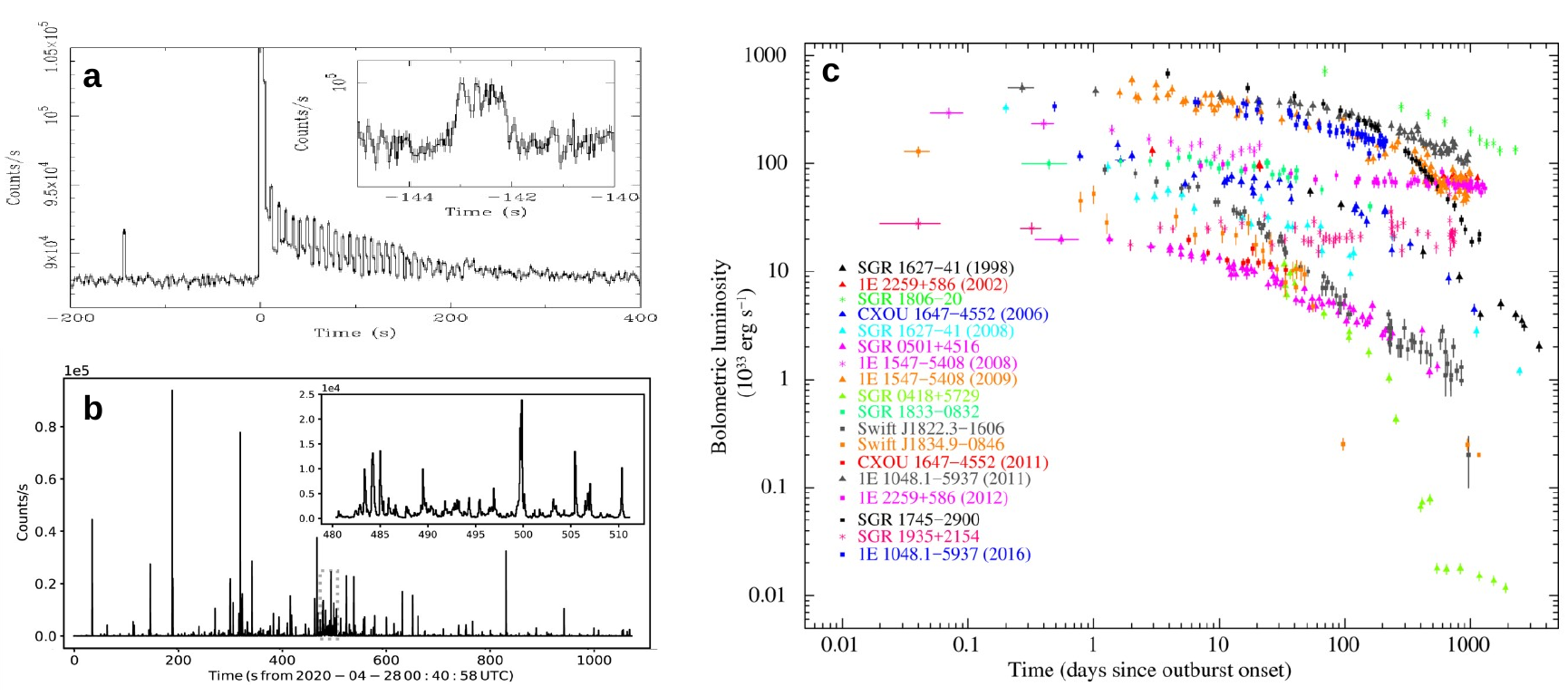}
\caption{Different types of magnetar transient activity. {\it Panel a}: Light curve of the most recent giant flare from SGR\,1806--20, recorded in 2004 with a 2.5\,s time resolution using {\it INTEGRAL} data. The initial spike reached an observed count rate $\gtrsim2\times10^6$\,counts\,s$^{-1}$. The inset shows the light curve of the precursor burst at full resolution (50\,ms). From \cite{mereghetti05}. {\it Panel b}: {\it NICER} 1--10\,keV light curve of the burst forest emitted from SGR\,1935+2154 in 2020 April. The light curve was extracted from the first $\sim$1200 seconds of observation, starting at 00:40:58 UTC on April 28, 2020. The inset zooms in on the area marked by a dotted gray box, highlighting the most intense bursting period. From \cite{Younes20}. {\it Panel c}: Temporal evolution of bolometric (0.01--100\,keV) luminosities for major outbursts up to the end of 2016 with good observational coverage. From \cite{Cotizelati18}.}
\label{fig:mag_transientactivity}
\end{figure}

Transient activity on different time scales is the distinctive trait of magnetars. This activity ranges from milliseconds to hundreds of seconds (e.g., giant flares and bursts) and also includes longer-lasting outbursts that can span months to years (see Figure\,\ref{fig:mag_transientactivity}). In the following, we discuss these phenomena in detail.

{\it Giant flares.} To date, only three giant flares from three different magnetars have been documented: in 1979 from SGR\,0526--66, in 1998 from SGR 1900+14, and in 2004 from SGR 1806--20 \citep{mazets79,Hurley99,palmer05}. The detection of only three such events over approximately 45 years implies these events are rare. The characteristics of the three giant flares closely resemble each other. They all commenced with a brief ($\sim0.1-0.2$\,s) spike of gamma-rays, reaching peak luminosities of $\gtrsim10^{44}-10^{45}$\,\lum\ for SGR 0526--66 and SGR 1900+14, and $\gtrsim10^{47}$\,\lum\ for SGR 1806--20. Following the initial flashes, they exhibited hard X-ray tails strongly modulated at the neutron star spin period and observed to decay within a few minutes (see Figure\,\ref{fig:mag_transientactivity}, Panel a). The energy released in the pulsating tails was intriguingly similar across all three events, approximately 10$^{44}$\,erg. According to the magnetar model, all the energy is discharged during the initial spike when a hot fireball is ejected. A portion of this energy is trapped within regions of closed magnetic field lines in the stellar magnetosphere and is converted into a photon-pair plasma, which cools down, producing the radiation observed in the oscillating tail. The fraction of trapped energy depends on the magnetic field strength. Hence, the similarity in the energy released during the three giant flares suggests that the three magnetars have magnetic fields of comparable magnitude.

At extra-galactic distances, only the brief initial hard spike of a giant flare would be observable, resembling a short gamma-ray burst. The short gamma-ray burst GRB\,051103 was historically the first extra-galactic giant flare candidate outside the Local Group, associated with the M81/M82 group of galaxies at a distance of $\approx$3.6\,Mpc. The most recent candidate is the short gamma-ray burst GRB\,231115A, which coincides with the nearby starburst galaxy M82 \citep{Mereghetti24}. The probability of this event being a giant flare in M82 rather than a short gamma-ray burst in the background is higher than for any other extra-galactic candidates, making it the most compelling case for a giant flare outside the Local Group. Moreover, the hard spectrum, short duration, and energy released during the burst align with the properties seen in the initial pulses of giant flares. However, it is noteworthy that the softer tail characterized by a periodic modulation induced by the neutron star rotation -- the signature of known giant flares -- remains unfortunately undetectable at extra-galactic distances using current instrumentation due to their faintness.


{\it Short bursts.} Short bursts are the hallmark of this class of isolated neutron stars, primarily aiding in the discovery of new magnetars and often signaling the onset of new outbursts. These bursts can occur sporadically or in clusters; however, it is impossible to predict when they will occur or which source is about to burst. Some magnetars may remain dormant for decades before suddenly entering an active phase, emitting hundreds of bursts within a few days. The duration of these bursts varies widely, ranging from millisecond to seconds, with peak luminosities typically falling within the range of $\sim10^{36}$ to 10$^{43}$\,\lum. The light curves of these bursts can show single or multiple peaks, with rise times generally faster than decay times. The brightest bursts, known as intermediate flares, can feature tails lasting for hours, resembling giant flares but potentially containing more energy than the initial spike. The energy ratios between the burst and tail can vary by an order of magnitude among different magnetars and even within bursts from the same source. Typically, magnetar bursts show no emission above $\sim$200\,keV. Their broadband spectra can be described by various models, including simple blackbody models, double blackbody models, optically thin thermal bremsstrahlung models, or Comptonized models (a power law with an exponential cutoff at higher energies). Often, more than one model can satisfactorily describe the burst spectral energy distribution, with a preference for single or double blackbody models. The blackbody temperature generally ranges between $\sim$2 and 12\,keV. When a second thermal component is required, a higher temperature ($\sim$12--13\,keV) blackbody with a smaller radius is typically observed alongside the lower temperature component.

SGR\,1935+2154 (SGR\,1935) stands out as one of the most prolific burst-producing magnetars \citep[see e.g.,][]{Lin20}. Since its discovery in 2014, it has emitted approximately 300 bursts, either as isolated events or during outbursts, including a burst forest in April 2020. The average burst energy increased over four episodes recorded between 2014 and 2016, indicating potentially more intense bursting activity with time. However, the burst energies during the 2019 and 2020 episodes (excluding the burst forest in 2020) suggested a plateau in the average burst energy curve. 
During its re-activation in 2020, SGR\,1935 experienced an extraordinary active phase. About six hours after the initial trigger, a burst forest occurred, with over 200 bursts in approximately 20 minutes, corresponding to a burst rate $>0.2$\,burst\,s$^{-1}$ (in stark contrast to the rate of 0.008\,burst\,s$^{-1}$ recorded three hours later, see Figure\,\ref{fig:mag_transientactivity}, Panel b; \citealt{Younes20}). 
However, what set this active phase apart was the detection of bright, millisecond-duration radio bursts \citep{Chime20,Bochenek20}, with properties reminiscent of fast radio bursts (FRBs), that is, powerful, millisecond-long bursts of radio emission from extra-galactic sources with unknown origins \citep[for a review see e.g.,][]{Petroff2022}. Some FRBs have been observed to emit more than one burst, indicating their progenitors are associated with non-cataclysmic events. Magnetars, due to their energetic nature, were speculated to be promising candidates for the central engine of repeating FRBs. The detection of such a burst from SGR\,1935 strengthened this paradigm. The energy released in the radio burst from SGR\,1935, $E_{\rm radio}\sim10^{34}-10^{35}$\,erg, was approximately three orders of magnitude greater than any radio pulse from the Crab pulsar, previously the source of the brightest Galactic radio bursts. Yet, it was 30 times less energetic than the weakest extra-galactic FRB observed to date. Notably, the radio burst coincided with a hard X-ray burst, marking the first time magnetar bursts have shown a bright radio counterpart (see e.g., \citealt{Mereghetti20}). While the X-ray burst spectrum was harder than any others from this source, its energy output, $E_{\rm X}\sim10^{39}$\,erg, fell within the expected range for magnetar bursts, resulting in $E_{\rm radio}/E_{\rm X}\sim10^{-5}$. Moreover, two FRB-like radio bursts were detected from the magnetar 1E\,1547.0--5408 during its 2009 outburst. One of them occurred $\approx$1\,s after an X-ray burst, yielding $E_{\rm radio}/E_{\rm X} \sim 10^{-9}$ \citep{Israel21}. These detections suggest that magnetar radio bursts exhibit a wide range of energies.


{\it Outbursts.}
The bursts often herald the start of an active phase, commonly referred to as an ``outburst''. During these episodes, the persistent X-ray luminosity suddenly increases by up to three orders of magnitude compared to its pre-outburst level. It then typically returns to its quiescent state over a period ranging from weeks to months, or even years (Figure\,\ref{fig:mag_transientactivity}, Panel c). Throughout an outburst, the X-ray spectrum initially hardens before gradually softening as the luminosity decreases. For example, in a scenario where a soft X-ray spectrum is modeled with an absorbed blackbody plus a power law, the hardening may correspond to an increase in the blackbody temperature along with a decrease in the photon index. If only the thermal component is needed for the quiescent spectrum, a second, hotter blackbody may appear, shrinking in size and decreasing in temperature as the outburst decays until it becomes undetectable. In some cases, a hard non-thermal tail is visible at the peak of the outburst, fading away faster than the softening of the 0.3–10\,keV spectrum.
Outbursts are thought to stem from heat accumulation within a confined region on the magnetar surface. However, the exact heating mechanism is still poorly understood. Magnetic stresses inject energy into the crust, which is then conducted to the surface, causing displacements and fractures. 
This energy injection mainly affects the outer crust, keeping most of the energy from being radiated as neutrinos (see \citealt{DeGrandis2022} and references therein). Additionally, a minimum energy threshold is needed to produce an observable effect. Energies below 10$^{40}$\,erg are too low to cause detectable increases in luminosity with current satellite sensitivity. Conversely, energies exceeding 10$^{43}$\,erg lead to a saturation effect, where additional energy does not alter the observable outcome. The surface photon luminosity reaches a maximum value of approximately 10$^{36}$\,\lum, because the crust becomes hot enough to release nearly all its energy as neutrinos before reaching the surface. Furthermore, the displacements in the crust induce a significant twist of the magnetic field lines in the magnetosphere, triggering additional surface heating due to currents flowing through the magnetic bundle and impacting upon the star surface \citep{Beloborodov09}. The twist must decay to supply its own currents. As the twist dissipates, the area affected by magnetospheric charges decreases, leading to a reduction in luminosity. Both of these mechanisms are likely at work during an outburst.

During an outburst, the morphology of pulse profiles can change significantly, ranging from broad shapes with one or two prominent peaks per cycle to more complex configurations as activity levels increase. Extreme and rapid variations in the spin-down torque are also frequently observed, which can significantly influence the long-term evolution of these sources. These variations are often caused by interactions between the magnetar crust and its magnetosphere, such as the injection of plasma into the magnetosphere and the subsequent changes in magnetic field configuration. Glitches, which are sudden spin-ups, frequently occur during outbursts and typically have frequency amplitudes in the range of $\delta \nu/\nu \sim 10^{-9}$ to 10$^{-5}$. 
Notable cases include 
1E\,2259$+$586 and SGR\,1935. 
For 1E\,2259+586, a candidate spin-down glitch in 2009 was accompanied by elevated flux levels. Two timing discontinuities were observed during an active phase in 2012. The first was an abrupt spin-down, or anti-glitch, and the second was attributed either to a spin-up glitch or another spin-down event, depending on the timing model. Two more timing irregularities were then detected over the following years. The first, occurring about five years after the 2012 anti-glitch, was a spin-up glitch without flux enhancement or changes in the spectral or pulse-profile shape. The second, in April 2019, was another anti-glitch, similar to the 2012 event, but without detectable changes in the pulse-profile shape, pulsed flux, or any bursts (see \citealt{Younes2020} and references therein). For SGR\,1935, an FRB-like burst was detected about three days after a significant spin-down glitch in October 2020, followed by an episode of pulsed radio emission lasting $\approx$2 weeks \citep{Younes2023,Zhu23}. Moreover, two glitches were observed during a subsequent outburst in October 2022, bracketing an FRB-like burst and ranking among the largest ever recorded from magnetars. Between the glitches, the magnetar exhibited a rapid spin-down phase, accompanied by an increase and subsequent decline in its persistent X-ray emission and burst rate \citep{Hu2024}. These examples suggest that glitches or anti-glitches may either not produce observable emission changes or be instead associated with significant radiative activity.

Pulsed radio emission has been detected in six magnetars during X-ray outbursts, in one case up to a frequency of 353\,GHz \citep{torne22}. In some instances, the radio emission persisted beyond the X-ray flux enhancement. The spectra of these radio-loud magnetars are generally flatter compared to RPPs: $S\propto\nu^{-0.5}$ for the former and $S\propto\nu^{-1.8}$ for the latter, where $S$ represents the flux density and $\nu$ denotes the frequency. The radio emission exhibits significant pulse-to-pulse variability, with pulse shapes varying considerably over minutes. Typically, single pulses consist of narrow, spiky sub-pulses with a high degree of linear polarization. Polarized quasi-periodic substructure has been identified in the emission of all well-studied radio-loud magnetars \citep{Kramer2024}. This substructure was detectable across all classes of radio-emitting neutron stars, suggesting a unifying feature in the radio emission mechanisms of these diverse objects. Recent observations of the magnetar XTE\,J1810--197 following its 2018 outburst have provided insights into these mechanisms \citep{Desvignes2024,Lower2024}. These observations revealed that the radio emission from XTE\,J1810--197 is composed of narrow, millisecond-duration subpulses with a highly polarized average pulse profile consisting of three distinct components. The polarization of these pulses exhibited remarkable temporal and frequency-dependent variations, including phase-dependent conversions of linear to circular polarization, attributed to birefringent and dispersive propagation effects within the magnetosphere. Additionally, systematic changes in the position angle (PA) of linear polarization over time were detected, including reversals in the sign of the PA gradient. 
These variations suggest that the magnetar experienced free precession after the outburst, while its magnetosphere gradually untwisted, with the precession being damped on a timescale of months. The changes in polarization and PA point to complex dynamics within the magnetar magnetosphere, possibly involving local twists in the magnetic field lines.

\subsection*{Low-magnetic field magnetars and high-magnetic field pulsars}

The discovery of three magnetars with dipolar magnetic fields within the range of those of ordinary radio pulsars, SGR\,0418+5729, Swift\,J1822.3--1606, and 3XMM\,J1852+0033, challenges the widespread belief that magnetars must have exceptionally powerful magnetic fields \citep[see e.g., for a review][]{Turolla13}. These sources are not markedly different from other members of the class, aside from the strength of their dipole magnetic fields inferred from the timing parameters, $B_{\rm dip}\sim(0.6-4)\times10^{13}$\,G, and their large characteristic ages, $\tau_{\rm c}>10^6$\,yr, which are two to three orders of magnitude higher than typical values for magnetars. These two properties suggest that these three objects might be old magnetars that have already experienced substantial field decay over their lifetimes. Additional features, including the limited number of observed bursts and the low luminosity during quiescence, support this interpretation. The critical question centers on whether a low dipolar magnetic field can cause crustal displacements, leading to bursting activity. Detailed models suggest that a local magnetic field of $\approx2\times10^{15}$\,G could result in crustal failures, while closer to the outer crust a magnetic field of $\approx10^{14}$\,G might be sufficient to break the crust. Nevertheless, the minimum requirement for this effect appears to be around $10^{14}$\,G. A crucial factor in producing bursts and outbursts is the internal toroidal component of the magnetic field, which is generally not measurable. For example, magneto-thermal evolutionary models suggest that the observed properties of SGR\,0418+5729 can be explained by assuming an initial external dipole field of $B_{\rm dip}\sim2\times10^{14}$\,G and an initial internal toroidal field of the order of 10$^{16}$\,G. This internal toroidal field is strong enough to cause crustal fractures at the present time. 

The presence of non-dipolar magnetic field components was observationally proven in the low-$B$ magnetar SGR\,0418+5729, through the detection of a phase-dependent absorption feature in the X-ray spectrum during an outburst \citep{Tiengo13}. The properties of this spectral line varied significantly with the star rotational phase: its energy ranged between $\sim$1 and 5\,keV over about one-fifth of a spin cycle. The most likely explanation for this variability is resonant cyclotron scattering. In this scenario, the phase dependence of the line central energy stems from the varying magnetic field strengths encountered by the charged particles as they interact with photons directed towards us while the neutron star rotates. The cyclotron energy for a particle of charge $q$ and mass $m$ can be expressed as 
\begin{equation}
E_{\rm cycl} = \frac{11.6}{1+z} \left( \frac{m_{\rm e}}{m} \right)B_{\rm 12}\,{\rm keV},  
\end{equation}
where $z=2GM_{\rm NS}/R_{\rm NS} c^2 \sim 0.4$ (assuming standard parameters for a neutron star: mass $M_{\rm NS}$=1.4$M_\odot$ and radius $R_{\rm NS}$=10\,km) and 1+$z$ is the gravitational redshift at the neutron star surface; $m_{\rm e}$ is the electron mass; $B_{\rm 12}$ is the magnetic field strength in units of 10$^{12}$\,G. If protons are the particles responsible for the scattering, the energy range of the feature requires a magnetic field $>2\times10^{14}$\,G, which is much higher than the surface dipolar component ($B_{\rm dip}\sim6\times10^{12}$\,G). According to the simplest model, scattering could take place within a localized magnetic loop near the magnetar surface. If this interpretation is correct, it supports the theoretical idea of a complex topology of magnetar magnetospheres with strong multi-polar components that can eventually break the crust and power outbursts. This feature was also observed in another low-$B$ magnetar, Swift\,J1822.3--1606 \citep{Rodriguez16}, again indicating the existence of confined magnetic coronal loops with magnetic fields of 10$^{14}$--10$^{15}$\,G (versus $B_{\rm dip}\sim3\times10^{13}$\,G). Two additional detections were reported in two XDINSs (see Sec.\,\ref{ins:xdin}). 

A handful of RPPs have periods and period derivatives similar to those of the magnetars, suggesting that they have comparable magnetic field strengths and occupy overlapping regions in the $P-\dot{P}$ diagram (Figure\,\ref{fig:ppdotdiagram}). Therefore, these high-$B$ RPPs might be expected to exhibit magnetar-like properties. In fact, two high-$B$ pulsars have shown magnetar outbursts. One of the youngest pulsars in the Galaxy, PSR\,J1846--0258, with a dipolar magnetic field strength of $B_{\rm dip}\sim5\times10^{13}$\,G, behaved as a typical RPP  for most of its observed lifespan. Its X-ray luminosity was below its spin-down luminosity, and it did not emit detectable radio signals, which might be due to unfavorable beaming conditions. The source experienced two magnetar outbursts, one in 2006 and another in 2020, both preceded by the emission of short magnetar-like X-ray bursts (see \citealt{Hu2023} and references therein). During these outbursts, the X-ray spectrum softened and required an additional component (a blackbody) along with the power law component that alone adequately describes the spectrum in its quiescent state. In the radio band, PSR\,J1119--6127 ($B_{\rm dip}\sim4\times10^{13}$\,G) is a stable radio pulsar. However, it experienced an uncommon glitch in 2007, followed by a brief change in the shape of the pulse profile. In 2016, it demonstrated its magnetically active nature  when two short hard X-ray bursts marked the onset of an outburst, increasing the flux by a factor of $>$160 (see \citealt{wang20} and references therein). These two sources can be considered transition objects between RPPs and magnetars, raising the possibility that more high-$B$ pulsars might be dormant magnetars and providing further evidence for a grand unification of the isolated neutron star population.

\section{X-ray dim isolated neutron stars} \label{ins:xdin}

Discovered in the nineties thanks to the high sensitivity of the {\it ROSAT} satellite in the soft X-rays, seven thermally emitting X-ray pulsars have been identified as a new class of isolated neutron stars, known as X-ray dim isolated neutron stars (XDINSs). These pulsars quickly earned the nickname ``The Magnificent Seven'' (for a review, see \citealt{Turolla09}). Each of these XDINSs has confirmed faint optical and/or ultraviolet counterparts, but multiple attempts to detect them at radio frequencies have been unsuccessful. XDINSs are among the closest known neutron stars, with distances estimated to be less than $\sim$500\,pc, based on modeling the distribution of the hydrogen column density $N_{\rm H}$. Parallax measurements are available for only two sources, RX\,J1856.5--3754 and RX\,J0720.4--3125, indicating distances of 123$^{+11}_{-15}$\,pc and 280$^{+210}_{-85}$\,pc, respectively. These measurements are in agreement with the distances estimated from the modelling of their spectra and $N_{\rm H}$ values. For all but one of these objects, the spin periods and their derivatives have been measured, ranging from $P\sim3-12$\,s and $\dot{P}\sim10^{-14}-10^{-13}$\,\ss, implying dipolar magnetic fields $B_{\rm dip}\sim(1-4)\times10^{13}$\,G and characteristic ages $\tau_{\rm c}\sim1-4$\,Myr. These properties place the XDINSs at the long-period end of the rotation-powered radio pulsar population and below the magnetars in the $P-\dot{P}$ diagram (see Figure\,\ref{fig:ppdotdiagram}). RX\,J1605.3+3249 is the only member without a coherent timing solution. A candidate spin period of 3.39\,s and period derivative of $1.6\times10^{-12}$\,\ss\ were suggested with a confidence level of 4$\sigma$ . This would correspond to $B_{\rm dip}\sim7.5\times10^{13}$\,G, the highest among the XDINSs if confirmed. However, a targeted campaign with {\it XMM-Newton} and {\it NICER} ruled out this candidate, setting stringent upper limits on the pulsed fraction of 1.3\% for periods above 0.15\,s (see \citealt{Malacaria19} and references therein). Recently, it was proposed that the most likely spin period of RX\,J0720.4--3125 is 16.78\,s, which is twice the previously reported period of 8.39\,s. A second peak at $P=16.78$\,s was identified in all {\it XMM-Newton} pointed observations across different energy bands. In some energy ranges, this peak was more significant than the one observed at $P=8.39$\,s. The light curves folded at the new claimed period exhibit a markedly double-peaked shape that varies with time and energy \citep{Hambaryan17}.  

XDINSs are observed to have nearly thermal spectra in soft X-rays (0.2--1.5\,keV), with temperatures between $\sim$50 and 100\,eV and low absorption column density values ($N_{\rm H}\sim10^{20}$\,\cm2). The soft X-ray emission is thought to arise from the neutron star surface. Consequently, these sources are ideal for testing atmosphere models and constraining the neutron star mass-to-radius ratio, ultimately helping to understand the equation of state of super-dense matter. For instance, spectral fitting of the phase-averaged spectra of RX\,J1308.6+2127 and RX\,J0720.4--3125 using a model that includes a condensed iron surface and a partially ionized hydrogen-thin atmosphere suggests a stiff equation of state. Assuming a standard neutron star mass of 1.4\,M$_{\odot}$, the mass-to-radius ratio ($M$/M$_{\odot}$)/($R$/km) is 0.087$\pm$0.004 for the former XDINS and 0.105$\pm$0.001 for the latter XDINS.

Although a blackbody model generally provides a good description of the soft X-ray spectrum of XDINSs, absorption features have been detected in five of them. These spectral lines share similar properties: they are centered at energies between $\sim$300\,eV and $\sim$800\,eV and are quite broad (widths$\sim$70--170\,keV) with equivalent widths of several tens of eV, $\sim$30--150\,eV. Moreover, the absorption features appear to vary with the spin phase. Their nature is unclear as yet, but three main explanations have been proposed for their origin. Firstly, these features could be caused by proton cyclotron resonances in a hot ionized layer near the surface, with line energies implying magnetic field strengths of the order of 10$^{13}$\,G, which is roughly consistent with values inferred from timing parameters. Another potential mechanism is atomic transitions in a magnetized atmosphere. The energies of the observed absorption features can be easily matched for hydrogen atmospheres, except for RX\,J2143.0+0654, where the line central energy of $\sim$0.7\,keV suggests a helium or heavier element atmosphere. Lastly, an inhomogeneous surface temperature distribution can also induce spectral distortions. Temperature anisotropies can be inferred from measurements of small blackbody emitting areas and are theoretically expected due to factors such as anisotropic thermal conductivity caused by a strong magnetic field. Predicting the temperature distribution on the neutron star surface is challenging due to many theoretical uncertainties, including the geometry of the magnetic field. Nevertheless, temperature inhomogeneities can significantly affect emission processes and should be considered alongside more sophisticated emission models, such as atmospheric or condensed surface models.

The two XDINSs with no detected spectral features are RX\,J1856.5--3754 and RX\,J0420.0--5022. Notably, they are the coldest in this class and the only ones for which a weak non-thermal component above $\sim$1.5\,keV has been reported (e.g., \citealt{degrandis2022b}). The cumulative X-ray spectrum of RX\,J1856.5--3754, extracted from 20 years of {\it XMM-Newton} observations with a total exposure time of 1.43\,Ms, showed a hard excess up to $\approx8$\,keV. An acceptable fit is obtained with a three-component model: a cold blackbody ($kT\sim60$\,eV, $R\sim5$\,km), which describes the soft part of the spectrum; a warmer and smaller blackbody emitting area ($kT\sim140$\,eV, $R\sim30$\,m) dominating up to $\approx2$\,keV and a non-thermal power law with a photon index $\Gamma\sim1.4$ emerging at higher energy. The luminosity of the non-thermal component corresponds to an efficiency of 10$^{-3}$, which is in line with what is observed in other rotation-powered X-ray pulsars with high spin-down power. Therefore, it is natural to assume a magnetospheric origin for this component. 
Overall, the spectral decomposition is most likely an approximation of a more complex thermal distribution on the neutron star surface resulting from its magneto-thermal evolution. For RX\,J0420.0--5022, the detection is not as strong as in the previous case due to the scarce amount of available data. An excess with respect to the one-blackbody model was found and was interpreted as either a second blackbody or a power law. These two XDINSs have the highest spin-down power among their siblings, making them the most promising candidates for producing non-thermal emission. Further investigations are necessary to determine whether these two sources are different from the other XDINSs or if all XDINSs have a non-thermal component in their emission.  

The discovery of phase-dependent absorption features in two low-magnetic-field magnetars, along with the belief that XDINSs are the descendants of magnetars according to magneto-thermal evolutionary  models, prompted the search for similar features in this class of sources. A phase-variable absorption feature with properties similar to those reported in the magnetars SGR\,0418+5729 and Swift\,J1822.3--1606 was found in a couple of XDINSs: RX\,J0720.4--3125 and RX J1308.6+2127  \citep{Borghese15,Borghese17}. Notably, the absorption features in both XDINSs are detected in only $\sim$20\% of their rotational cycle and have remained stable over the time span covered by the observations, $\sim$12\,yr for RX\,J0720.4--3125 and $\sim$7\,yr for RX\,J1308.6+2127. The central energy of the feature is $\sim$0.75\,keV, with an equivalent width of $\sim$15--30\,eV. Given the similarities with SGR\,0418+5729, this feature might be explained by the same physical mechanism: proton cyclotron resonant scattering. In this scenario, the sharp variability with phase is attributed to small-scale ($\sim$100\,m) magnetic structures near the neutron star surface. The inferred magnetic field in these loops is about five times stronger than the dipolar component for both sources. This discovery supports the evolutionary connection between XDINSs and magnetars. It also provides observational evidence that the magnetic fields of highly magnetized isolated neutron stars are more complex than simple dipoles, featuring small-scale deviations such as localized high magnetic field bundles.

All the XDINSs have confirmed faint optical and ultraviolet counterparts through {\it Hubble Space Telescope} observations. The optical/ultraviolet flux exceeds the value derived from the extrapolation of the best-fit model of the X-ray spectra, a phenomenon referred to as optical excess. Most XDINSs exhibit an optical excess value (defined as the ratio between the measured flux and the flux predicted by model extrapolation) between 5 and 12, except for one source, RX\,J2143.0+0654, which exceeds the X-ray extrapolation by a factor of more than 50 \citep{Kaplan11}. The origin of this optical excess remains unclear. If the X-ray and optical radiation originate from different regions on the neutron star surface, variations in the optical excess should correlate with changes in the X-ray pulsed fraction: a small X-ray hot spot would result in both a significant optical excess and a high pulsed fraction. However, no such correlations have been observed, suggesting that this explanation may be incomplete. Alternative scenarios suggest the optical excess could be due to magnetospheric emission, atmospheric effects, or the presence of a nebula. Extended near-infrared emission, $\sim$0.8\,arcsec in size, was discovered around RX\,J0806.4--4123. This emission showed an inhomogeneous brightness distribution, consisting of a brighter core and an elongated halo. The near-infrared flux exceeded the expected levels based on the extrapolation of the optical/ultraviolet spectrum. This emission could be interpreted as originating from a pulsar wind nebula, fueled by low-energy shocked pulsar wind particles, or from a disk with a favorable viewing geometry \citep{Posselt18b}. The spectral signatures of these two scenarios differ from each other and can be investigated with high-resolution observations, such as those from the {\it James Webb Space Telescope}. Additionally, deep near-infrared surveys are needed to determine whether such extended emission is common among isolated neutron stars or if RX\,J0806.4--4123 is an exception.

In recent years, the launch of the eROSITA instrument \citep{predehl21} on board the Spectrum-Roentgen-Gamma satellite has significantly improved the prospects of expanding the XDINS sample. eROSITA offers a much improved sensitivity at soft X-ray energies and better positional accuracy than its predecessor, {\it ROSAT}. Indeed, eROSITA, along with subsequent observations with {\it XMM-Newton} and the {\it VLT}, has already led to the discovery of a new XDINS, eRASSU\,J131716.9$-$402647 \citep{kurpas24}. The X-ray emission from this source was strongly pulsed, with a period of 12.8\,s, showing a double-humped pulse profile and a pulsed fraction increasing with energy. The spectrum was best described by a purely thermal continuum, modeled either as a blackbody with a temperature of $\sim$95\,eV or a magnetized neutron star atmosphere at $\sim$35\,eV. In addition, a broad absorption feature at 260\,eV and a narrow one at 590\,eV were detected. Deep optical observations did not detect a counterpart, implying an X-ray-to-optical flux ratio of $\gtrsim$10$^4$. The discovery of eRASSU\,J131716.9$-$402647 was part of a larger systematic search within the eROSITA western Galactic hemisphere down to an X-ray flux limit of 10$^{-13}$\,\flux. This search identified 33 new XDINS candidates, which were divided into two groups based on their spectral properties: 13 with soft X-ray emission ($kT\sim40-100$\,eV), aligning with known XDINS characteristics; and 20 with harder X-ray emission, which could be younger, hotter neutron stars or other types of high-energy sources. 
Future observations in the X-ray and optical bands will help confirm their nature as isolated neutron stars. Searches for periodic emission and for features in the X-ray spectrum may provide constraints on the magnetic field near the neutron star surface.

\section{Central compact objects} 
\label{ins:cco}

The enigmatic family of central compact objects (CCOs) consists of about a dozen sources\footnote{See the complete catalog at \url{http://www.iasf-milano.inaf.it/~deluca/cco/main.htm}.}. These sources are located near the center of young (0.3–7\,kyr) supernova remnants and emit X-rays without counterparts at any other wavelength \citep[for a review see e.g.,][]{Deluca17}. These objects are observed as soft ($\sim$0.2--5\,keV) X-ray sources with a thermal spectrum that is well described by the sum of two blackbodies with temperatures $kT\sim0.2-0.5$\,keV and small emitting radii spanning from 0.1 to a few km. Their emission is usually steady with a luminosity of the order of 10$^{33}$\,\lum\ and pulsations are detected in only three CCOs. While only a handful of CCOs are currently known, their locations in young supernova remnants suggest they may represent a significant fraction of all neutron star births. The designation {\it CCO} was originally introduced by \cite{Pavlov00} to denote the point-like source situated at the center of the supernova remnant Cassiopeia A. Since then, it has served as the label for objects sharing the aforementioned traits, albeit potentially misleadingly. For example, while the Crab pulsar resides at the center of its nebula and is a compact object, it does not meet the criteria for classification as a CCO because its emission extends beyond X-rays to other wavelengths.

\begin{figure}[t]
\centering
\includegraphics[width=0.95\textwidth, trim={2cm 5cm 1.5cm 0cm},clip=true]{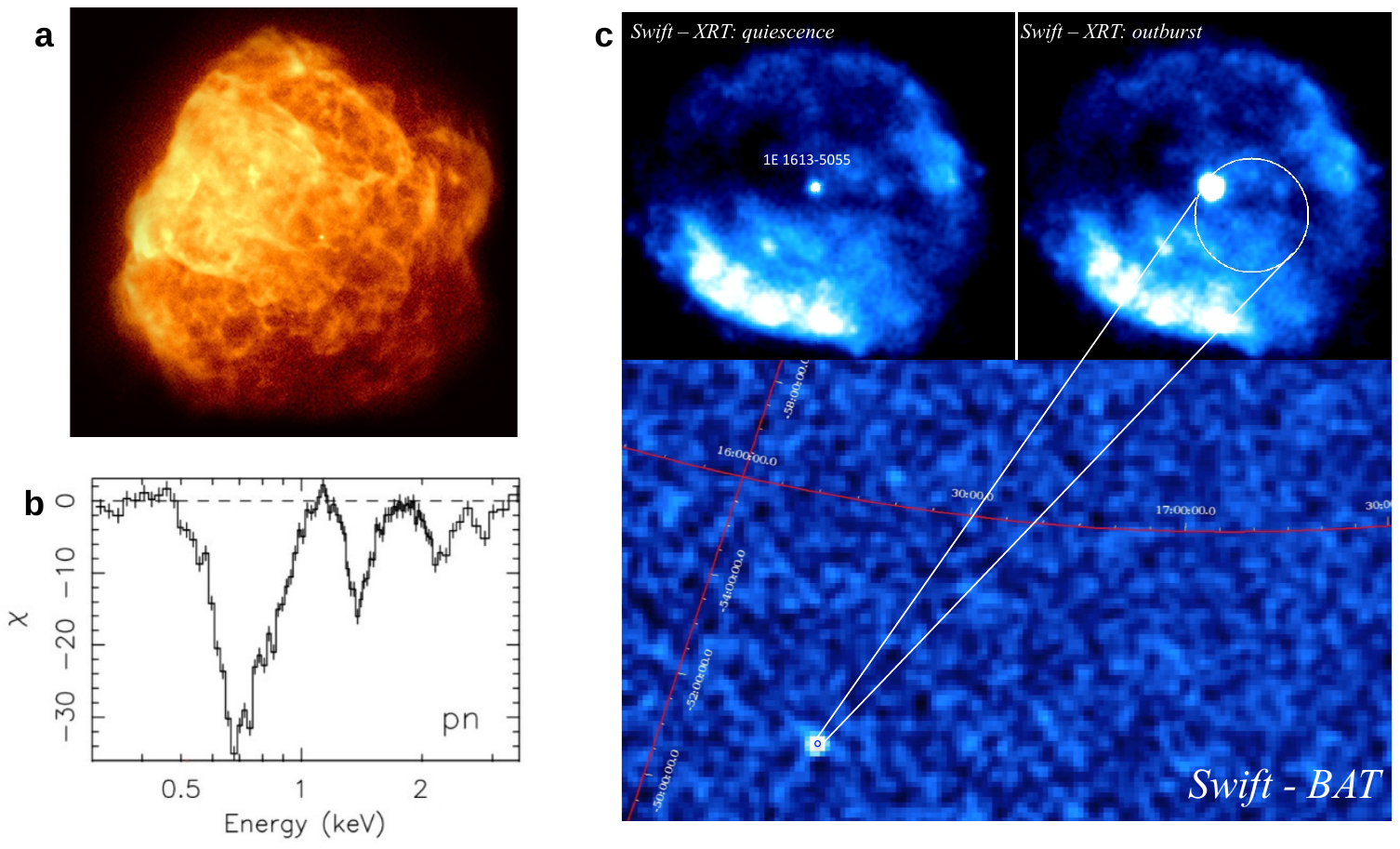}
\caption{{\it Panel a}: Image in X-rays of the supernova remnant Puppis\,A with its central compact object PSR\,J0821--4300, taken with the Follow-up X-ray Telescope onboard of {\it Einstein Probe} launched on 9 January 2024. Image credit: EPSC, NAO/CAS; DSS; ESO. {\it Panel b}: Residuals in units of sigma obtained from a comparison between the data with the best-fit thermal continuum model (i.e., the sum of two blackbodies) for 1E\,1207.4--5209. The presence of four absorption spectral features at $\sim$0.7, 1.4, 2.1, and 2.8 keV is evident. From \cite{Deluca04}. {\it Panel c}: {\it Swift}/Burst Alert Telescope 15--150\,keV image of the burst detected in June 2016  from the direction of the supernova remnant RCW\,103 (bottom). Two {\it Swift}/X-ray Telescope co-added 1--10\,keV images of
RCW\,103 during the quiescence of 1E\,161348--5055 (from 2011 April 18 to 2016 May 16; exposure time of 66\,ks; top left) and in outburst (from 2016 June 22 to July 20; exposure time of 67\,ks; top right). The white circle marks the positional accuracy of the detected burst, which has a radius of 1.5\,arcmin. From \cite{Rea16}.}
\label{fig:ccos}
\end{figure}

The confirmation that CCOs are indeed neutron stars was strengthened by the detection of pulsations from three objects: PSR\,J1852+0040 in Kes\,79 ($P\sim105$\,ms), PSR\,J0821--4300 in Puppis\,A ($P\sim112$\,ms; see Figure\,\ref{fig:ccos}, Panel a), and 1E\,1207.4--5209 in G\,296.5+10.0 ($P\sim424$\,ms). Extensive observational campaigns were required to measure their period derivatives, which turn out to be very small ($\dot{P}\sim10^{-18}-10^{-17}$\,\ss). The spin parameters unveiled several key insights: {\it (i)}  their spin-down luminosity is about 10 times lower than their X-ray luminosity; {\it (ii)} the characteristic age $\tau_{\rm c}$ is 4--5 orders of magnitude greater than the age of the host supernova remnants, suggesting that CCOs were either born spinning with periods close to their current ones or underwent unusual magnetic field evolution; {\it (iii)} the inferred dipolar magnetic field $B_{\rm dip}\sim10^{10}-10^{11}$\,G is the weakest among all known young isolated neutron stars. Due to their weak magnetic fields, CCOs were awarded the epithet of {\it anti-magnetars}: neutron stars born with low magnetic fields that have not been significantly amplified by dynamo effect, likely due to their slow rotation at birth. Nonetheless, the anti-magnetar scenario fails to address various observational characteristics.

For all three pulsating CCOs, the X-ray spectral energy distribution reveals a continuum emission of thermal origin, described by the sum of two blackbodies with small emitting radii. These tiny hot spots are inconsistent with the inferred magnetic field values, as surface temperature anisotropies are typically associated with stronger magnetic fields. The thermal emission of PSR\,J1852+0040 is strongly modulated, with a pulsed fraction of $\sim$64\%. Its pulse profile features a single broad pulse, whose shape is energy-independent. In contrast, PSR\,J0821--4300 displays a sinusoidal pulse profile with a 180$^{\circ}$ phase reversal at $\sim$1.2\,keV, where the hot and warm blackbodies switch dominance. This suggests the presence of antipodal hot spots with different temperatures and areas, which are difficult to explain if the magnetic field is weak. The X-ray spectrum of 1E\,1207.4--5209 displays four absorption features centered at $\sim$0.7, 1.4, 2.1, and 2.8\,keV, aso shown in Figure\,\ref{fig:ccos}, Panel b. Since these spectral lines are harmonically spaced, the most natural explanation is cyclotron absorption involving one fundamental frequency and three harmonics. This interpretation provides a direct measurement of the neutron star magnetic field of $\sim8\times10^{10}$\,G. This result is consistent with the dipolar component of the magnetic field inferred from the spin-down parameters, $B_{\rm dip} \simeq 9.8 \times10^{10}$\,G. 

Thanks to a 20-year observing campaign, timing irregularities were found in the ephemeris of 1E\,1207.4-5209, identified as either two small glitches or extreme timing noise \citep{Gotthelf20}. This was the first discovery of such an event in a CCO and in an isolated neutron star with such a small period derivative. Notably, no changes in the spectrum or in the energy centroids of the spectral features were observed alongside these timing anomalies. This suggests that the surface magnetic field strength remained constant before and after the event. Therefore, internal effects, such as the motion of an internal magnetic field comparable in strength to those of canonical young pulsars with similar timing behaviors, are favored as the cause of the glitches or timing noise. This finding provides additional support for the leading theory of CCOs: they are born with a normal-strength magnetic field, which is later buried by fallback debris following the supernova explosion (the so-called {\it hidden magnetic field} scenario). 
Several models suggest that accreting a mass of $\sim10^{-4}-10^{-2}$,M$_{\odot}$ can push canonical ($\sim10^{12}$\,G) or strong ($\sim10^{14}$\,G) magnetic fields of a newborn neutron star into its crust. This results in an external magnetic field that is weaker than the hidden internal field, which might re-emerge on time scales of $10^{3}-10^{5}$\,yr after accretion ends. Following this phase, the surface magnetic field is restored to a value close to its birth strength. The thermal evolution during the re-emergence phase can lead to varying degrees of anisotropy in the surface temperature, accounting for the significant pulsed fraction of PSR\,J1852+00409 and the antipodal hot spots of PSR \,J0821--4300. Moreover, the slow re-emergence of the buried field may contribute to triggering glitches or excess timing noise, as observed in 1E\,1207.4--5209. 

In addition to the three pulsating CCOs, this group includes approximately ten other sources. These sources exhibit similar emission properties, showing no detected pulsations (with deep upper limits in a few cases) and no long-term variability, despite the lack of multi-epoch campaigns in most cases. This suggests that these sources are isolated neutron stars akin to the three CCO pulsars, possibly with even smaller dipole fields.
The CCO at the center of the supernova remnant Cassiopeia\,A is one of the most popular CCOs. Discovered during the first-light observations of {\it Chandra}, it is among the youngest neutron stars in the Milky Way, with the age of the supernova remnant estimated to be $\sim$350\,yr. Various models provide a satisfactory description of its X-ray spectrum. These include blackbody models and both magnetic and non-magnetic hydrogen atmosphere models, all of which suggest that the emission originates from small hot spots, similar to those found in other CCOs. However, timing investigations have not detected pulsations, with a pulsed fraction limit of $\sim$12\%, suggesting either a uniform temperature distribution on the neutron star surface or an unfavorable orientation of the observer's line of sight relative to the spin and magnetic dipole axes. A viable option to reconcile these conflicting results is offered by low magnetized ($B_{\rm dip}<10^{11}$\,G) carbon atmosphere models. These models provide a good spectral fit with emission from the entire neutron star surface, which does not necessarily vary with rotation. Directly measuring the cooling of an isolated neutron star could have significant theoretical implications for understanding their internal composition and structure, since applying cooling models can help constrain neutrino emission mechanisms and envelope compositions. Interestingly, multi-epoch {\it Chandra} observations indicated a long-term evolution in the luminosity and spectral shape of this source, compatible with a drop in the temperature of $\sim4\%$ over about $\sim10$\,yr (2000--2009). However, these observations were affected by several instrumental effects that caused time-dependent spectral distortions. To minimize these effects, additional observations with {\it Chandra} were conducted in 2013, 2015, and 2020 \citep{Posselt22}. An apparent increase in the cooling rate was reported between 2015 and 2020. The authors note that these changes might be due to systematic effects, possibly from imperfect calibration of the increasing contamination in the optical-blocking filter.

Carbon atmosphere models with low magnetic fields effectively describe the X-ray spectrum not only of the CCO in Cassiopeia\,A, but also of a few more sources. These models assume a homogeneous temperature distribution across the entire neutron star surface, which naturally explains the absence of detected pulsations within current limits. Additionally, they aid in investigating the equation of state for dense matter. A robust distance measurement allows for meaningful constraints on the mass and radius of isolated neutron stars. For instance, for the CCO within the supernova remnant HESS\,J1731--347 (also known as G353.6--0.7), the X-ray spectrum was modeled using state-of-the-art carbon atmosphere model. Taking advantage of Gaia data to obtain an accurate distance estimation (2.5(3)\,kpc), the mass and radius of the neutron star were estimated equal to 0.8$\pm$0.2\,M$_\odot$ and 10.4$_{-0.8}^{+0.9}$\,km, respectively \citep{Doroshenko2022}. These measurements make it the lightest known neutron star or possibly a `strange star' with an exotic equation of state.


\subsection{A disguised CCO}

The discovery of 1E\,161348--5055 (1E\,1613) dates back to 1980 when the {\it Einstein} observatory detected a point-like source at the center of the young ($\sim$2\,kyr) supernova remnant RCW\,103. Although considered one of the CCO prototypes, 1E\,1613 exhibits several unique characteristics that set it apart from all other CCOs and isolated neutron stars \citep{Deluca06}. Firstly, unlike other CCOs, 1E\,1613 shows significant flux variations on a monthly to yearly timescale, including an outburst in 1999 when its flux increased by a factor of $\sim$100. Secondly, it exhibits an unusual periodicity of 6.67\,hr with strong quasi-sinusoidal modulation. This 6.67-hr periodicity was detected in all sufficiently long observations, but the pulse profile changed with the source flux state: it is sinusoidal during low states (observed soft X-ray flux $\sim10^{-12}$\,\flux) and more complex during high states ($\sim10^{-11}$\,\flux).
These distinctive properties have led to the proposal that 1E\,1613 might be the first low-mass X-ray binary in a supernova remnant with an orbital period of 6.67\,hr, or a young compact object with a long spin period of 6.67\,hr. The latter interpretation leans toward a magnetar interpretation, which can naturally account for the flux variability and changes in the pulse profile.

A breakthrough occurred in June 2016, when a short magnetar-like burst of hard X-rays from the direction of RCW\,103 triggered all-sky X-ray monitors \citep[see Figure\,\ref{fig:ccos}, Panel c;][]{Rea16}. Follow-up observations revealed that 1E\,1613 was experiencing an outburst akin to that of a magnetar: its luminosity was about 100 times higher than the quiescent level that had persisted for years and was observed up to one month before. Additionally, the light curve displayed two peaks per cycle, which was a dramatic change from the sinusoidal shape seen during its quiescent state. For the first time, hard X-ray, non-thermal emission up to $\sim$30\,keV was observed and modeled with a power law, while the soft X-ray spectrum was well fit with a double-blackbody model. Moreover, an infrared counterpart was identified, which had not been detected in previous observations. The X-ray to infrared fluence ratio of $\sim10^5$ was consistent with measurements for magnetars. However, it is still debated whether the infrared emission originates from the neutron star magnetosphere or a fallback disk. While these results align with the hallmark features of magnetars, the periodicity is puzzling, being slower by three orders of magnitude than any other known magnetar candidate. 
An efficient braking mechanism must be invoked to slow down the source from a fast birth period ($<0.5$\,s) to the long current value in $\sim2$\,kyr. Most models consider a propeller interaction with a fallback disk, which could offer an additional spin-down torque along with that caused by dipole radiation. 
For example, a remnant disk of $\sim10^{-9}$\,M$_{\odot}$ could slow down 1E\,1613 from milliseconds down to hours in $\sim1-3$\,kyr, if the dipolar magnetic field is $B_{\rm dip}\sim5\times10^{15}$\,G, a value slightly higher than the magnetic fields of known magnetars \citep{Ho17}.

\section{Final remarks} 

The study of isolated neutron stars has uncovered a wide variety of these fascinating objects, each distinguished by unique observational properties and emission mechanisms. Extensive observations and theoretical modeling have significantly advanced our understanding of RPPs, magnetars, XDINSs and CCOs. Each class offers valuable insights into the underlying physics of neutron stars, including their magnetic fields, rotational dynamics, emission processes, and evolutionary paths.

Looking ahead, the \emph{Square Kilometre Array} (SKA) will revolutionize radio astronomy with unprecedented sensitivity, enabling the discovery of many more pulsars and detailed studies of their properties. Large-scale radio surveys with the SKA and other telescopes will be critical in uncovering new, peculiar neutron stars, such as long-period radio sources, and understanding their full population diversity. High-energy telescopes like the recently launched \emph{Einstein-Probe}, and the upcoming \emph{NewAthena} and the \emph{Cherenkov Telescope Array Observatory}, will provide valuable data to probe the (very) high-energy emission mechanisms at work in these sources. These missions will also help detect and study transient phenomena such as magnetar bursts and outbursts in greater detail than ever before.

Overall, the study of isolated neutron stars is poised for significant advancements, with each discovery offering new opportunities for exploration and a deeper understanding of these enigmatic objects.

\begin{ack}[Acknowledgments]
 AB is supported by a L'Or\'eal--UNESCO For Women in Science Fellowship (XXI Italian edition). FCZ is supported by a Ram\'on y Cajal fellowship (grant agreement RYC2021-030888-I) and acknowledges financial support from the Spanish program Unidad de Excelencia María de Maeztu CEX2020-001058-M and the Italian National Institute for Astrophysics (INAF) Research Grant ``Uncovering the optical beat of the fastest magnetised neutron stars'' (FANS).
\end{ack}


\bibliographystyle{Harvard}
\bibliography{INS}

\begin{thebibliography*}{80}
\providecommand{\bibtype}[1]{}
\providecommand{\natexlab}[1]{#1}
{\catcode`\|=0\catcode`\#=12\catcode`\@=11\catcode`\\=12
|immediate|write|@auxout{\expandafter\ifx\csname
  natexlab\endcsname\relax\gdef\natexlab#1{#1}\fi}}
\renewcommand{\url}[1]{{\tt #1}}
\providecommand{\urlprefix}{URL }
\expandafter\ifx\csname urlstyle\endcsname\relax
  \providecommand{\doi}[1]{doi:\discretionary{}{}{}#1}\else
  \providecommand{\doi}{doi:\discretionary{}{}{}\begingroup
  \urlstyle{rm}\Url}\fi
\providecommand{\bibinfo}[2]{#2}
\providecommand{\eprint}[2][]{\url{#2}}

\bibtype{Article}%
\bibitem[{Alpar} et al.(1982)]{Alpar1982}
\bibinfo{author}{{Alpar} MA}, \bibinfo{author}{{Cheng} AF},
  \bibinfo{author}{{Ruderman} MA} and  \bibinfo{author}{{Shaham} J}
  (\bibinfo{year}{1982}).
\bibinfo{title}{{A new class of radio pulsars}}.
\bibinfo{journal}{{\em Nature}} \bibinfo{volume}{300} (\bibinfo{number}{5894}):
  \bibinfo{pages}{728--730}.

\bibtype{Article}%
\bibitem[Andersen et al.(2020)]{Chime20}
\bibinfo{author}{Andersen BC}, \bibinfo{author}{Bandura KM},
  \bibinfo{author}{Bhardwaj M}, \bibinfo{author}{Bij A}, \bibinfo{author}{Boyce
  MM}, \bibinfo{author}{Boyle PJ}, \bibinfo{author}{Brar C},
  \bibinfo{author}{Cassanelli T}, \bibinfo{author}{Chawla P},
  \bibinfo{author}{Chen T}, \bibinfo{author}{Cliche JF}, \bibinfo{author}{Cook
  A}, \bibinfo{author}{Cubranic D}, \bibinfo{author}{Curtin AP},
  \bibinfo{author}{Denman NT}, \bibinfo{author}{Dobbs M}, \bibinfo{author}{Dong
  FQ}, \bibinfo{author}{Fandino M}, \bibinfo{author}{Fonseca E},
  \bibinfo{author}{Gaensler BM}, \bibinfo{author}{Giri U},
  \bibinfo{author}{Good DC}, \bibinfo{author}{Halpern M}, \bibinfo{author}{Hill
  AS}, \bibinfo{author}{Hinshaw GF}, \bibinfo{author}{H{\"o}fer C},
  \bibinfo{author}{Josephy A}, \bibinfo{author}{Kania JW},
  \bibinfo{author}{Kaspi VM}, \bibinfo{author}{Landecker TL},
  \bibinfo{author}{Leung C}, \bibinfo{author}{Li DZ}, \bibinfo{author}{Lin HH},
  \bibinfo{author}{Masui KW}, \bibinfo{author}{Mckinven R},
  \bibinfo{author}{Mena-Parra J}, \bibinfo{author}{Merryfield M},
  \bibinfo{author}{Meyers BW}, \bibinfo{author}{Michilli D},
  \bibinfo{author}{Milutinovic N}, \bibinfo{author}{Mirhosseini A},
  \bibinfo{author}{M{\"u}nchmeyer M}, \bibinfo{author}{Naidu A},
  \bibinfo{author}{Newburgh LB}, \bibinfo{author}{Ng C}, \bibinfo{author}{Patel
  C}, \bibinfo{author}{Pen UL}, \bibinfo{author}{Pinsonneault-Marotte T},
  \bibinfo{author}{Pleunis Z}, \bibinfo{author}{Quine BM},
  \bibinfo{author}{Rafiei-Ravandi M}, \bibinfo{author}{Rahman M},
  \bibinfo{author}{Ransom SM}, \bibinfo{author}{Renard A},
  \bibinfo{author}{Sanghavi P}, \bibinfo{author}{Scholz P},
  \bibinfo{author}{Shaw JR}, \bibinfo{author}{Shin K}, \bibinfo{author}{Siegel
  SR}, \bibinfo{author}{Singh S}, \bibinfo{author}{Smegal RJ},
  \bibinfo{author}{Smith KM}, \bibinfo{author}{Stairs IH}, \bibinfo{author}{Tan
  CM}, \bibinfo{author}{Tendulkar SP}, \bibinfo{author}{Tretyakov I},
  \bibinfo{author}{Vanderlinde K}, \bibinfo{author}{Wang H},
  \bibinfo{author}{Wulf D}, \bibinfo{author}{Zwaniga AV} and
  \bibinfo{author}{Collaboration TC} (\bibinfo{year}{2020}).
\bibinfo{title}{A bright millisecond-duration radio burst from a galactic
  magnetar}.
\bibinfo{journal}{{\em Nature}} \bibinfo{volume}{587} (\bibinfo{number}{7832}):
  \bibinfo{pages}{54--58}.

\bibtype{Article}%
\bibitem[{Baade} and {Zwicky}(1934)]{Baade1934}
\bibinfo{author}{{Baade} W} and  \bibinfo{author}{{Zwicky} F}
  (\bibinfo{year}{1934}).
\bibinfo{title}{{Cosmic Rays from Super-novae}}.
\bibinfo{journal}{{\em Proceedings of the National Academy of Science}}
  \bibinfo{volume}{20} (\bibinfo{number}{5}): \bibinfo{pages}{259--263}.

\bibtype{Article}%
\bibitem[{Beloborodov}(2009)]{Beloborodov09}
\bibinfo{author}{{Beloborodov} AM} (\bibinfo{year}{2009}).
\bibinfo{title}{{Untwisting Magnetospheres of Neutron Stars}}.
\bibinfo{journal}{{\em The Astrophysical Journal}} \bibinfo{volume}{703}
  (\bibinfo{number}{1}): \bibinfo{pages}{1044--1060}.

\bibtype{Article}%
\bibitem[{Beloborodov} and {Thompson}(2007)]{beloborodov07}
\bibinfo{author}{{Beloborodov} AM} and  \bibinfo{author}{{Thompson} C}
  (\bibinfo{year}{2007}).
\bibinfo{title}{{Corona of Magnetars}}.
\bibinfo{journal}{{\em The Astrophysical Journal}} \bibinfo{volume}{657}:
  \bibinfo{pages}{967--993}.

\bibtype{Book}%
\bibitem[Bhattacharyya et al.(2022)]{msp2022}
\bibinfo{editor}{Bhattacharyya S}, \bibinfo{editor}{Papitto A} and
  \bibinfo{editor}{Bhattacharya D}, (Eds.)  (\bibinfo{year}{2022}).
\bibinfo{title}{{Millisecond Pulsars}}, \bibinfo{series}{Astrophysics and Space
  Science Library}, \bibinfo{volume}{465}.

\bibtype{Article}%
\bibitem[Bochenek et al.(2020)]{Bochenek20}
\bibinfo{author}{Bochenek CD}, \bibinfo{author}{Ravi V}, \bibinfo{author}{Belov
  KV}, \bibinfo{author}{Hallinan G}, \bibinfo{author}{Kocz J},
  \bibinfo{author}{Kulkarni SR} and  \bibinfo{author}{McKenna DL}
  (\bibinfo{year}{2020}).
\bibinfo{title}{A fast radio burst associated with a galactic magnetar}.
\bibinfo{journal}{{\em Nature}} \bibinfo{volume}{587} (\bibinfo{number}{7832}):
  \bibinfo{pages}{59--62}.

\bibtype{incollection}%
\bibitem[{Borghese} and {Esposito}(2023)]{borghese23}
\bibinfo{author}{{Borghese} A} and  \bibinfo{author}{{Esposito} P}
  (\bibinfo{year}{2023}), \bibinfo{title}{{Isolated Neutron Stars}},
  \bibinfo{booktitle}{Handbook of X-ray and Gamma-ray Astrophysics}, pp.
  \bibinfo{pages}{146}.

\bibtype{Article}%
\bibitem[{Borghese} et al.(2015)]{Borghese15}
\bibinfo{author}{{Borghese} A}, \bibinfo{author}{{Rea} N},
  \bibinfo{author}{{Coti Zelati} F}, \bibinfo{author}{{Tiengo} A} and
  \bibinfo{author}{{Turolla} R} (\bibinfo{year}{2015}).
\bibinfo{title}{{Discovery of a Strongly Phase-variable Spectral Feature in the
  Isolated Neutron Star RX J0720.4-3125}}.
\bibinfo{journal}{{\em The Astrophysical Journal Letters}}
  \bibinfo{volume}{807} (\bibinfo{number}{1}), \bibinfo{eid}{L20}.

\bibtype{Article}%
\bibitem[{Borghese} et al.(2017)]{Borghese17}
\bibinfo{author}{{Borghese} A}, \bibinfo{author}{{Rea} N},
  \bibinfo{author}{{Coti Zelati} F}, \bibinfo{author}{{Tiengo} A},
  \bibinfo{author}{{Turolla} R} and  \bibinfo{author}{{Zane} S}
  (\bibinfo{year}{2017}).
\bibinfo{title}{{Narrow phase-dependent features in X-ray dim isolated neutron
  stars: a new detection and upper limits}}.
\bibinfo{journal}{{\em Monthly Notices of the Royal Astronomical Society}}
  \bibinfo{volume}{468} (\bibinfo{number}{3}): \bibinfo{pages}{2975--2983}.

\bibtype{Article}%
\bibitem[{Caleb} et al.(2022)]{Caleb2022}
\bibinfo{author}{{Caleb} M}, \bibinfo{author}{{Heywood} I},
  \bibinfo{author}{{Rajwade} K}, \bibinfo{author}{{Malenta} M},
  \bibinfo{author}{{Stappers} BW}, \bibinfo{author}{{Barr} E},
  \bibinfo{author}{{Chen} W}, \bibinfo{author}{{Morello} V},
  \bibinfo{author}{{Sanidas} S}, \bibinfo{author}{{van den Eijnden} J},
  \bibinfo{author}{{Kramer} M}, \bibinfo{author}{{Buckley} D},
  \bibinfo{author}{{Brink} J}, \bibinfo{author}{{Motta} SE},
  \bibinfo{author}{{Woudt} P}, \bibinfo{author}{{Weltevrede} P},
  \bibinfo{author}{{Jankowski} F}, \bibinfo{author}{{Surnis} M},
  \bibinfo{author}{{Buchner} S}, \bibinfo{author}{{Bezuidenhout} MC},
  \bibinfo{author}{{Driessen} LN} and  \bibinfo{author}{{Fender} R}
  (\bibinfo{year}{2022}).
\bibinfo{title}{{Discovery of a radio-emitting neutron star with an ultra-long
  spin period of 76 s}}.
\bibinfo{journal}{{\em Nature Astronomy}} \bibinfo{volume}{6}:
  \bibinfo{pages}{828--836}.

\bibtype{Article}%
\bibitem[{Chrimes} et al.(2022)]{Chrimes2022}
\bibinfo{author}{{Chrimes} AA}, \bibinfo{author}{{Levan} AJ},
  \bibinfo{author}{{Fruchter} AS}, \bibinfo{author}{{Groot} PJ},
  \bibinfo{author}{{Kouveliotou} C}, \bibinfo{author}{{Lyman} JD},
  \bibinfo{author}{{Tanvir} NR} and  \bibinfo{author}{{Wiersema} K}
  (\bibinfo{year}{2022}).
\bibinfo{title}{{New candidates for magnetar counterparts from a deep search
  with the Hubble Space Telescope}}.
\bibinfo{journal}{{\em Monthly Notices of the Royal Astronomical Society}}
  \bibinfo{volume}{512} (\bibinfo{number}{4}): \bibinfo{pages}{6093--6103}.

\bibtype{Article}%
\bibitem[{Coti Zelati} et al.(2018)]{Cotizelati18}
\bibinfo{author}{{Coti Zelati} F}, \bibinfo{author}{{Rea} N},
  \bibinfo{author}{{Pons} JA}, \bibinfo{author}{{Campana} S} and
  \bibinfo{author}{{Esposito} P} (\bibinfo{year}{2018}).
\bibinfo{title}{{Systematic study of magnetar outbursts}}.
\bibinfo{journal}{{\em Monthly Notices of the Royal Astronomical Society}}
  \bibinfo{volume}{474}: \bibinfo{pages}{961--1017}.

\bibtype{Article}%
\bibitem[{De Grandis} et al.(2022{\natexlab{a}})]{degrandis2022b}
\bibinfo{author}{{De Grandis} D}, \bibinfo{author}{{Rigoselli} M},
  \bibinfo{author}{{Mereghetti} S}, \bibinfo{author}{{Younes} G},
  \bibinfo{author}{{Pizzochero} P}, \bibinfo{author}{{Taverna} R},
  \bibinfo{author}{{Tiengo} A}, \bibinfo{author}{{Turolla} R} and
  \bibinfo{author}{{Zane} S} (\bibinfo{year}{2022}{\natexlab{a}}).
\bibinfo{title}{{Two decades of X-ray observations of the isolated neutron star
  RX J1856.5 - 3754: detection of thermal and non-thermal hard X-rays and
  refined spin-down measurement}}.
\bibinfo{journal}{{\em Monthly Notices of the Royal Astronomical Society}}
  \bibinfo{volume}{516} (\bibinfo{number}{4}): \bibinfo{pages}{4932--4941}.

\bibtype{Article}%
\bibitem[{De Grandis} et al.(2022{\natexlab{b}})]{DeGrandis2022}
\bibinfo{author}{{De Grandis} D}, \bibinfo{author}{{Turolla} R},
  \bibinfo{author}{{Taverna} R}, \bibinfo{author}{{Lucchetta} E},
  \bibinfo{author}{{Wood} TS} and  \bibinfo{author}{{Zane} S}
  (\bibinfo{year}{2022}{\natexlab{b}}).
\bibinfo{title}{{Three-dimensional Magnetothermal Simulations of Magnetar
  Outbursts}}.
\bibinfo{journal}{{\em The Astrophysical Journal}} \bibinfo{volume}{936}
  (\bibinfo{number}{2}), \bibinfo{eid}{99}.

\bibtype{Inproceedings}%
\bibitem[{De Luca}(2017)]{Deluca17}
\bibinfo{author}{{De Luca} A} (\bibinfo{year}{2017}), \bibinfo{title}{{Central
  compact objects in supernova remnants}}, \bibinfo{booktitle}{Journal of
  Physics Conference Series}, \bibinfo{series}{Journal of Physics Conference
  Series}, \bibinfo{volume}{932}, pp. \bibinfo{pages}{012006}.

\bibtype{Article}%
\bibitem[{De Luca} et al.(2004)]{Deluca04}
\bibinfo{author}{{De Luca} A}, \bibinfo{author}{{Mereghetti} S},
  \bibinfo{author}{{Caraveo} PA}, \bibinfo{author}{{Moroni} M},
  \bibinfo{author}{{Mignani} RP} and  \bibinfo{author}{{Bignami} GF}
  (\bibinfo{year}{2004}).
\bibinfo{title}{{XMM-Newton and VLT observations of the isolated neutron star
  1E 1207.4-5209}}.
\bibinfo{journal}{{\em Astronomy and Astrophysics}} \bibinfo{volume}{418}:
  \bibinfo{pages}{625--637}.

\bibtype{Article}%
\bibitem[{De Luca} et al.(2006)]{Deluca06}
\bibinfo{author}{{De Luca} A}, \bibinfo{author}{{Caraveo} PA},
  \bibinfo{author}{{Mereghetti} S}, \bibinfo{author}{{Tiengo} A} and
  \bibinfo{author}{{Bignami} GF} (\bibinfo{year}{2006}).
\bibinfo{title}{{A Long-Period, Violently Variable X-ray Source in a Young
  Supernova Remnant}}.
\bibinfo{journal}{{\em Science}} \bibinfo{volume}{313}
  (\bibinfo{number}{5788}): \bibinfo{pages}{814--817}.

\bibtype{Article}%
\bibitem[{Desvignes} et al.(2024)]{Desvignes2024}
\bibinfo{author}{{Desvignes} G}, \bibinfo{author}{{Weltevrede} P},
  \bibinfo{author}{{Gao} Y}, \bibinfo{author}{{Jones} DI},
  \bibinfo{author}{{Kramer} M}, \bibinfo{author}{{Caleb} M},
  \bibinfo{author}{{Karuppusamy} R}, \bibinfo{author}{{Levin} L},
  \bibinfo{author}{{Liu} K}, \bibinfo{author}{{Lyne} AG},
  \bibinfo{author}{{Shao} L}, \bibinfo{author}{{Stappers} B} and
  \bibinfo{author}{{P{\'e}tri} J} (\bibinfo{year}{2024}).
\bibinfo{title}{{A freely precessing magnetar following an X-ray outburst}}.
\bibinfo{journal}{{\em Nature Astronomy}} .

\bibtype{Article}%
\bibitem[{Dhillon} et al.(2009)]{Dhillon09}
\bibinfo{author}{{Dhillon} VS}, \bibinfo{author}{{Marsh} TR},
  \bibinfo{author}{{Littlefair} SP}, \bibinfo{author}{{Copperwheat} CM},
  \bibinfo{author}{{Kerry} P}, \bibinfo{author}{{Dib} R},
  \bibinfo{author}{{Durant} M}, \bibinfo{author}{{Kaspi} VM},
  \bibinfo{author}{{Mignani} RP} and  \bibinfo{author}{{Shearer} A}
  (\bibinfo{year}{2009}).
\bibinfo{title}{{Optical pulsations from the anomalous X-ray pulsar
  1E1048.1-5937}}.
\bibinfo{journal}{{\em Monthly Notices of the Royal Astronomical Society}}
  \bibinfo{volume}{394} (\bibinfo{number}{1}): \bibinfo{pages}{L112--L116}.

\bibtype{Article}%
\bibitem[{Doroshenko} et al.(2022)]{Doroshenko2022}
\bibinfo{author}{{Doroshenko} V}, \bibinfo{author}{{Suleimanov} V},
  \bibinfo{author}{{P{\"u}hlhofer} G} and  \bibinfo{author}{{Santangelo} A}
  (\bibinfo{year}{2022}).
\bibinfo{title}{{A strangely light neutron star within a supernova remnant}}.
\bibinfo{journal}{{\em Nature Astronomy}} \bibinfo{volume}{6}:
  \bibinfo{pages}{1444--1451}.

\bibtype{Article}%
\bibitem[{Duncan} and {Thompson}(1992)]{Duncan92}
\bibinfo{author}{{Duncan} RC} and  \bibinfo{author}{{Thompson} C}
  (\bibinfo{year}{1992}).
\bibinfo{title}{{Formation of Very Strongly Magnetized Neutron Stars:
  Implications for Gamma-Ray Bursts}}.
\bibinfo{journal}{{\em The Astrophysical Journal Letters}}
  \bibinfo{volume}{392}: \bibinfo{pages}{L9}.

\bibtype{Article}%
\bibitem[{Enoto} et al.(2017)]{enoto17}
\bibinfo{author}{{Enoto} T}, \bibinfo{author}{{Shibata} S},
  \bibinfo{author}{{Kitaguchi} T}, \bibinfo{author}{{Suwa} Y},
  \bibinfo{author}{{Uchide} T}, \bibinfo{author}{{Nishioka} H},
  \bibinfo{author}{{Kisaka} S}, \bibinfo{author}{{Nakano} T},
  \bibinfo{author}{{Murakami} H} and  \bibinfo{author}{{Makishima} K}
  (\bibinfo{year}{2017}).
\bibinfo{title}{{Magnetar Broadband X-Ray Spectra Correlated with Magnetic
  Fields: Suzaku Archive of SGRs and AXPs Combined with NuSTAR, Swift, and
  RXTE}}.
\bibinfo{journal}{{\em The Astrophysical Journal Supplement Series}}
  \bibinfo{volume}{231}, \bibinfo{eid}{8}.

\bibtype{Article}%
\bibitem[{Giacconi} et al.(1962)]{Giacconi1962}
\bibinfo{author}{{Giacconi} R}, \bibinfo{author}{{Gursky} H},
  \bibinfo{author}{{Paolini} FR} and  \bibinfo{author}{{Rossi} BB}
  (\bibinfo{year}{1962}).
\bibinfo{title}{{Evidence for x Rays From Sources Outside the Solar System}}.
\bibinfo{journal}{{\em Physical Review Letters}} \bibinfo{volume}{9}
  (\bibinfo{number}{11}): \bibinfo{pages}{439--443}.

\bibtype{Article}%
\bibitem[{Gold}(1968)]{gold68}
\bibinfo{author}{{Gold} T} (\bibinfo{year}{1968}).
\bibinfo{title}{{Rotating Neutron Stars as the Origin of the Pulsating Radio
  Sources}}.
\bibinfo{journal}{{\em Nature}} \bibinfo{volume}{218} (\bibinfo{number}{5143}):
  \bibinfo{pages}{731--732}.

\bibtype{Article}%
\bibitem[{Gotthelf} and {Halpern}(2020)]{Gotthelf20}
\bibinfo{author}{{Gotthelf} EV} and  \bibinfo{author}{{Halpern} JP}
  (\bibinfo{year}{2020}).
\bibinfo{title}{{The Timing Behavior of the Central Compact Object Pulsar 1E
  1207.4-5209}}.
\bibinfo{journal}{{\em The Astrophysical Journal}} \bibinfo{volume}{900}
  (\bibinfo{number}{2}), \bibinfo{eid}{159}.

\bibtype{Article}%
\bibitem[{G{\"o}tz} et al.(2006)]{Gotz06}
\bibinfo{author}{{G{\"o}tz} D}, \bibinfo{author}{{Mereghetti} S},
  \bibinfo{author}{{Tiengo} A} and  \bibinfo{author}{{Esposito} P}
  (\bibinfo{year}{2006}).
\bibinfo{title}{{Magnetars as persistent hard X-ray sources: INTEGRAL discovery
  of a hard tail in SGR 1900+14}}.
\bibinfo{journal}{{\em Astronomy and Astrophysics}} \bibinfo{volume}{449}
  (\bibinfo{number}{2}): \bibinfo{pages}{L31--L34}.

\bibtype{Article}%
\bibitem[{Hambaryan} et al.(2017)]{Hambaryan17}
\bibinfo{author}{{Hambaryan} V}, \bibinfo{author}{{Suleimanov} V},
  \bibinfo{author}{{Haberl} F}, \bibinfo{author}{{Schwope} AD},
  \bibinfo{author}{{Neuh{\"a}user} R}, \bibinfo{author}{{Hohle} M} and
  \bibinfo{author}{{Werner} K} (\bibinfo{year}{2017}).
\bibinfo{title}{{The compactness of the isolated neutron star RX
  J0720.4-3125}}.
\bibinfo{journal}{{\em Astronomy and Astrophysics}} \bibinfo{volume}{601},
  \bibinfo{eid}{A108}.

\bibtype{Article}%
\bibitem[{Harding} et al.(2021)]{Harding2021}
\bibinfo{author}{{Harding} AK}, \bibinfo{author}{{Venter} C} and
  \bibinfo{author}{{Kalapotharakos} C} (\bibinfo{year}{2021}).
\bibinfo{title}{{Very-high-energy Emission from Pulsars}}.
\bibinfo{journal}{{\em The Astrophysical Journal}} \bibinfo{volume}{923}
  (\bibinfo{number}{2}), \bibinfo{eid}{194}.

\bibtype{Article}%
\bibitem[{Hare} et al.(2024)]{Hare2024}
\bibinfo{author}{{Hare} J}, \bibinfo{author}{{Pavlov} GG},
  \bibinfo{author}{{Posselt} B}, \bibinfo{author}{{Kargaltsev} O},
  \bibinfo{author}{{Temim} T} and  \bibinfo{author}{{Chen} S}
  (\bibinfo{year}{2024}).
\bibinfo{title}{{Probing the spectrum of the magnetar 4U 0142+61 with JWST}}.
\bibinfo{journal}{{\em arXiv e-prints}} , \bibinfo{eid}{arXiv:2405.03947}.

\bibtype{Article}%
\bibitem[{Hewish} et al.(1968)]{Hewish68}
\bibinfo{author}{{Hewish} A}, \bibinfo{author}{{Bell} SJ},
  \bibinfo{author}{{Pilkington} JDH}, \bibinfo{author}{{Scott} PF} and
  \bibinfo{author}{{Collins} RA} (\bibinfo{year}{1968}).
\bibinfo{title}{{Observation of a Rapidly Pulsating Radio Source}}.
\bibinfo{journal}{{\em Nature}} \bibinfo{volume}{217} (\bibinfo{number}{5130}):
  \bibinfo{pages}{709--713}.

\bibtype{Article}%
\bibitem[{Ho} and {Andersson}(2017)]{Ho17}
\bibinfo{author}{{Ho} WCG} and  \bibinfo{author}{{Andersson} N}
  (\bibinfo{year}{2017}).
\bibinfo{title}{{Ejector and propeller spin-down: how might a superluminous
  supernova millisecond magnetar become the 6.67 h pulsar in RCW 103}}.
\bibinfo{journal}{{\em Monthly Notices of the Royal Astronomical Society:
  Letters}} \bibinfo{volume}{464} (\bibinfo{number}{1}):
  \bibinfo{pages}{L65--L69}.

\bibtype{Article}%
\bibitem[{Hu} et al.(2023)]{Hu2023}
\bibinfo{author}{{Hu} CP}, \bibinfo{author}{{Kuiper} L},
  \bibinfo{author}{{Harding} AK}, \bibinfo{author}{{Younes} G},
  \bibinfo{author}{{Blumer} H}, \bibinfo{author}{{Ho} WCG},
  \bibinfo{author}{{Enoto} T}, \bibinfo{author}{{Espinoza} CM} and
  \bibinfo{author}{{Gendreau} K} (\bibinfo{year}{2023}).
\bibinfo{title}{{A NICER View on the 2020 Magnetar-like Outburst of PSR
  J1846-0258}}.
\bibinfo{journal}{{\em The Astrophysical Journal}} \bibinfo{volume}{952}
  (\bibinfo{number}{2}), \bibinfo{eid}{120}.

\bibtype{Article}%
\bibitem[{Hu} et al.(2024)]{Hu2024}
\bibinfo{author}{{Hu} CP}, \bibinfo{author}{{Narita} T},
  \bibinfo{author}{{Enoto} T}, \bibinfo{author}{{Younes} G},
  \bibinfo{author}{{Wadiasingh} Z}, \bibinfo{author}{{Baring} MG},
  \bibinfo{author}{{Ho} WCG}, \bibinfo{author}{{Guillot} S},
  \bibinfo{author}{{Ray} PS}, \bibinfo{author}{{G{\"u}ver} T},
  \bibinfo{author}{{Rajwade} K}, \bibinfo{author}{{Arzoumanian} Z},
  \bibinfo{author}{{Kouveliotou} C}, \bibinfo{author}{{Harding} AK} and
  \bibinfo{author}{{Gendreau} KC} (\bibinfo{year}{2024}).
\bibinfo{title}{{Rapid spin changes around a magnetar fast radio burst}}.
\bibinfo{journal}{{\em Nature}} \bibinfo{volume}{626} (\bibinfo{number}{7999}):
  \bibinfo{pages}{500--504}.

\bibtype{Article}%
\bibitem[{Hurley} et al.(1999)]{Hurley99}
\bibinfo{author}{{Hurley} K}, \bibinfo{author}{{Cline} T},
  \bibinfo{author}{{Mazets} E}, \bibinfo{author}{{Barthelmy} S},
  \bibinfo{author}{{Butterworth} P}, \bibinfo{author}{{Marshall} F},
  \bibinfo{author}{{Palmer} D}, \bibinfo{author}{{Aptekar} R},
  \bibinfo{author}{{Golenetskii} S}, \bibinfo{author}{{Il'Inskii} V},
  \bibinfo{author}{{Frederiks} D}, \bibinfo{author}{{McTiernan} J},
  \bibinfo{author}{{Gold} R} and  \bibinfo{author}{{Trombka} J}
  (\bibinfo{year}{1999}).
\bibinfo{title}{{A giant periodic flare from the soft {\ensuremath{\gamma}}-ray
  repeater SGR1900+14}}.
\bibinfo{journal}{{\em Nature}} \bibinfo{volume}{397} (\bibinfo{number}{6714}):
  \bibinfo{pages}{41--43}.

\bibtype{Article}%
\bibitem[{Hurley-Walker} et al.(2022)]{HurleyWalker2022}
\bibinfo{author}{{Hurley-Walker} N}, \bibinfo{author}{{Zhang} X},
  \bibinfo{author}{{Bahramian} A}, \bibinfo{author}{{McSweeney} SJ},
  \bibinfo{author}{{O'Doherty} TN}, \bibinfo{author}{{Hancock} PJ},
  \bibinfo{author}{{Morgan} JS}, \bibinfo{author}{{Anderson} GE},
  \bibinfo{author}{{Heald} GH} and  \bibinfo{author}{{Galvin} TJ}
  (\bibinfo{year}{2022}).
\bibinfo{title}{{A radio transient with unusually slow periodic emission}}.
\bibinfo{journal}{{\em Nature}} \bibinfo{volume}{601} (\bibinfo{number}{7894}):
  \bibinfo{pages}{526--530}.

\bibtype{Article}%
\bibitem[{Hurley-Walker} et al.(2023)]{HurleyWalker2023}
\bibinfo{author}{{Hurley-Walker} N}, \bibinfo{author}{{Rea} N},
  \bibinfo{author}{{McSweeney} SJ}, \bibinfo{author}{{Meyers} BW},
  \bibinfo{author}{{Lenc} E}, \bibinfo{author}{{Heywood} I},
  \bibinfo{author}{{Hyman} SD}, \bibinfo{author}{{Men} YP},
  \bibinfo{author}{{Clarke} TE}, \bibinfo{author}{{Coti Zelati} F},
  \bibinfo{author}{{Price} DC}, \bibinfo{author}{{Horv{\'a}th} C},
  \bibinfo{author}{{Galvin} TJ}, \bibinfo{author}{{Anderson} GE},
  \bibinfo{author}{{Bahramian} A}, \bibinfo{author}{{Barr} ED},
  \bibinfo{author}{{Bhat} NDR}, \bibinfo{author}{{Caleb} M},
  \bibinfo{author}{{Dall'Ora} M}, \bibinfo{author}{{de Martino} D},
  \bibinfo{author}{{Giacintucci} S}, \bibinfo{author}{{Morgan} JS},
  \bibinfo{author}{{Rajwade} KM}, \bibinfo{author}{{Stappers} B} and
  \bibinfo{author}{{Williams} A} (\bibinfo{year}{2023}).
\bibinfo{title}{{A long-period radio transient active for three decades}}.
\bibinfo{journal}{{\em Nature}} \bibinfo{volume}{619} (\bibinfo{number}{7970}):
  \bibinfo{pages}{487--490}.

\bibtype{Article}%
\bibitem[{Israel} et al.(2021)]{Israel21}
\bibinfo{author}{{Israel} GL}, \bibinfo{author}{{Burgay} M},
  \bibinfo{author}{{Rea} N}, \bibinfo{author}{{Esposito} P},
  \bibinfo{author}{{Possenti} A}, \bibinfo{author}{{Dall'Osso} S},
  \bibinfo{author}{{Stella} L}, \bibinfo{author}{{Pilia} M},
  \bibinfo{author}{{Tiengo} A}, \bibinfo{author}{{Ridnaia} A},
  \bibinfo{author}{{Lien} AY}, \bibinfo{author}{{Frederiks} DD} and
  \bibinfo{author}{{Bernardini} F} (\bibinfo{year}{2021}).
\bibinfo{title}{{X-Ray and Radio Bursts from the Magnetar 1E 1547.0-5408}}.
\bibinfo{journal}{{\em The Astrophysical Journal}} \bibinfo{volume}{907}
  (\bibinfo{number}{1}), \bibinfo{eid}{7}.

\bibtype{Article}%
\bibitem[{Kaplan} et al.(2011)]{Kaplan11}
\bibinfo{author}{{Kaplan} DL}, \bibinfo{author}{{Kamble} A},
  \bibinfo{author}{{van Kerkwijk} MH} and  \bibinfo{author}{{Ho} WCG}
  (\bibinfo{year}{2011}).
\bibinfo{title}{{New Optical/Ultraviolet Counterparts and the Spectral Energy
  Distributions of Nearby, Thermally Emitting, Isolated Neutron Stars}}.
\bibinfo{journal}{{\em The Astrophysical Journal}} \bibinfo{volume}{736}
  (\bibinfo{number}{2}), \bibinfo{eid}{117}.

\bibtype{Article}%
\bibitem[{Keane} and {McLaughlin}(2011)]{Keane2011}
\bibinfo{author}{{Keane} EF} and  \bibinfo{author}{{McLaughlin} MA}
  (\bibinfo{year}{2011}).
\bibinfo{title}{{Rotating radio transients}}.
\bibinfo{journal}{{\em Bulletin of the Astronomical Society of India}}
  \bibinfo{volume}{39} (\bibinfo{number}{3}): \bibinfo{pages}{333--352}.

\bibtype{Article}%
\bibitem[{Kouveliotou} et al.(1993)]{kouveliotou93}
\bibinfo{author}{{Kouveliotou} C}, \bibinfo{author}{{Fishman} GJ},
  \bibinfo{author}{{Meegan} CA}, \bibinfo{author}{{Paciesas} WS},
  \bibinfo{author}{{Wilson} RB}, \bibinfo{author}{{van Paradijs} J},
  \bibinfo{author}{{Preece} RD}, \bibinfo{author}{{Briggs} MS},
  \bibinfo{author}{{Pendleton} GN} and  \bibinfo{author}{{Brock} MN}
  (\bibinfo{year}{1993}).
\bibinfo{title}{{Recurrent burst activity from the soft gamma-ray repeater
  SGR\,1900+14}}.
\bibinfo{journal}{{\em Nature}} \bibinfo{volume}{362}:
  \bibinfo{pages}{728--730}.

\bibtype{Article}%
\bibitem[{Kramer} et al.(2024)]{Kramer2024}
\bibinfo{author}{{Kramer} M}, \bibinfo{author}{{Liu} K},
  \bibinfo{author}{{Desvignes} G}, \bibinfo{author}{{Karuppusamy} R} and
  \bibinfo{author}{{Stappers} BW} (\bibinfo{year}{2024}).
\bibinfo{title}{{Quasi-periodic sub-pulse structure as a unifying feature for
  radio-emitting neutron stars}}.
\bibinfo{journal}{{\em Nature Astronomy}} \bibinfo{volume}{8}:
  \bibinfo{pages}{230--240}.

\bibtype{Article}%
\bibitem[{Kurpas} et al.(2024)]{kurpas24}
\bibinfo{author}{{Kurpas} J}, \bibinfo{author}{{Schwope} AD},
  \bibinfo{author}{{Pires} AM} and  \bibinfo{author}{{Haberl} F}
  (\bibinfo{year}{2024}).
\bibinfo{title}{{Detection of pulsed X-ray emission from the isolated neutron
  star candidate eRASSU J131716.9-402647}}.
\bibinfo{journal}{{\em Astronomy and Astrophysics}} \bibinfo{volume}{683},
  \bibinfo{eid}{A164}.

\bibtype{Article}%
\bibitem[{Kuzmin}(2007)]{Kuzmin2007}
\bibinfo{author}{{Kuzmin} AD} (\bibinfo{year}{2007}).
\bibinfo{title}{{Giant pulses of pulsar radio emission}}.
\bibinfo{journal}{{\em Astrophysics and Space Science}} \bibinfo{volume}{308}
  (\bibinfo{number}{1-4}): \bibinfo{pages}{563--567}.

\bibtype{Article}%
\bibitem[{Lin} et al.(2020)]{Lin20}
\bibinfo{author}{{Lin} L}, \bibinfo{author}{{G{\"o}{\u{g}}{\"u}{\c{s}}} E},
  \bibinfo{author}{{Roberts} OJ}, \bibinfo{author}{{Kouveliotou} C},
  \bibinfo{author}{{Kaneko} Y}, \bibinfo{author}{{van der Horst} AJ} and
  \bibinfo{author}{{Younes} G} (\bibinfo{year}{2020}).
\bibinfo{title}{{Burst Properties of the Most Recurring Transient Magnetar SGR
  J1935+2154}}.
\bibinfo{journal}{{\em The Astrophysical Journal}} \bibinfo{volume}{893}
  (\bibinfo{number}{2}), \bibinfo{eid}{156}.

\bibtype{Article}%
\bibitem[{Lower} et al.(2024)]{Lower2024}
\bibinfo{author}{{Lower} ME}, \bibinfo{author}{{Johnston} S},
  \bibinfo{author}{{Lyutikov} M}, \bibinfo{author}{{Melrose} DB},
  \bibinfo{author}{{Shannon} RM}, \bibinfo{author}{{Weltevrede} P},
  \bibinfo{author}{{Caleb} M}, \bibinfo{author}{{Camilo} F},
  \bibinfo{author}{{Cameron} AD}, \bibinfo{author}{{Dai} S},
  \bibinfo{author}{{Hobbs} G}, \bibinfo{author}{{Li} D},
  \bibinfo{author}{{Rajwade} KM}, \bibinfo{author}{{Reynolds} JE},
  \bibinfo{author}{{Sarkissian} JM} and  \bibinfo{author}{{Stappers} BW}
  (\bibinfo{year}{2024}).
\bibinfo{title}{{Linear to circular conversion in the polarized radio emission
  of a magnetar}}.
\bibinfo{journal}{{\em Nature Astronomy}} .

\bibtype{Article}%
\bibitem[{Malacaria} et al.(2019)]{Malacaria19}
\bibinfo{author}{{Malacaria} C}, \bibinfo{author}{{Bogdanov} S},
  \bibinfo{author}{{Ho} WCG}, \bibinfo{author}{{Enoto} T},
  \bibinfo{author}{{Ray} PS}, \bibinfo{author}{{Arzoumanian} Z},
  \bibinfo{author}{{Cazeau} T}, \bibinfo{author}{{Gendreau} KC},
  \bibinfo{author}{{Guillot} S}, \bibinfo{author}{{G{\"u}ver} T},
  \bibinfo{author}{{Jaisawal} GK}, \bibinfo{author}{{Wolff} MT},
  \bibinfo{author}{{NICER Magnetar}} and  \bibinfo{author}{{Magnetospheres
  Team}} (\bibinfo{year}{2019}).
\bibinfo{title}{{A Joint NICER and XMM-Newton View of the ``Magnificent''
  Thermally Emitting X-Ray Isolated Neutron Star RX J1605.3+3249}}.
\bibinfo{journal}{{\em The Astrophysical Journal}} \bibinfo{volume}{880}
  (\bibinfo{number}{2}), \bibinfo{eid}{74}.

\bibtype{Article}%
\bibitem[{Manchester} et al.(2005)]{Manchester05}
\bibinfo{author}{{Manchester} RN}, \bibinfo{author}{{Hobbs} GB},
  \bibinfo{author}{{Teoh} A} and  \bibinfo{author}{{Hobbs} M}
  (\bibinfo{year}{2005}).
\bibinfo{title}{{The Australia Telescope National Facility Pulsar Catalogue}}.
\bibinfo{journal}{{\em The Astronomical Journal}} \bibinfo{volume}{129}
  (\bibinfo{number}{4}): \bibinfo{pages}{1993--2006}.

\bibtype{Article}%
\bibitem[{Mazets} et al.(1979)]{mazets79}
\bibinfo{author}{{Mazets} EP}, \bibinfo{author}{{Golentskii} SV},
  \bibinfo{author}{{Ilinskii} VN}, \bibinfo{author}{{Aptekar} RL} and
  \bibinfo{author}{{Guryan} IA} (\bibinfo{year}{1979}).
\bibinfo{title}{{Observations of a flaring X-ray pulsar in Dorado}}.
\bibinfo{journal}{{\em Nature}} \bibinfo{volume}{282}:
  \bibinfo{pages}{587--589}.

\bibtype{Article}%
\bibitem[{Mereghetti} et al.(2005)]{mereghetti05}
\bibinfo{author}{{Mereghetti} S}, \bibinfo{author}{{G{\"o}tz} D},
  \bibinfo{author}{{von Kienlin} A}, \bibinfo{author}{{Rau} A},
  \bibinfo{author}{{Lichti} G}, \bibinfo{author}{{Weidenspointner} G} and
  \bibinfo{author}{{Jean} P} (\bibinfo{year}{2005}).
\bibinfo{title}{{The First Giant Flare from SGR 1806-20: Observations Using the
  Anticoincidence Shield of the Spectrometer on INTEGRAL}}.
\bibinfo{journal}{{\em The Astrophysical Journal}} \bibinfo{volume}{624}:
  \bibinfo{pages}{L105--L108}.

\bibtype{Article}%
\bibitem[{Mereghetti} et al.(2020)]{Mereghetti20}
\bibinfo{author}{{Mereghetti} S}, \bibinfo{author}{{Savchenko} V},
  \bibinfo{author}{{Ferrigno} C}, \bibinfo{author}{{G{\"o}tz} D},
  \bibinfo{author}{{Rigoselli} M}, \bibinfo{author}{{Tiengo} A},
  \bibinfo{author}{{Bazzano} A}, \bibinfo{author}{{Bozzo} E},
  \bibinfo{author}{{Coleiro} A}, \bibinfo{author}{{Courvoisier} TJL},
  \bibinfo{author}{{Doyle} M}, \bibinfo{author}{{Goldwurm} A},
  \bibinfo{author}{{Hanlon} L}, \bibinfo{author}{{Jourdain} E},
  \bibinfo{author}{{von Kienlin} A}, \bibinfo{author}{{Lutovinov} A},
  \bibinfo{author}{{Martin-Carrillo} A}, \bibinfo{author}{{Molkov} S},
  \bibinfo{author}{{Natalucci} L}, \bibinfo{author}{{Onori} F},
  \bibinfo{author}{{Panessa} F}, \bibinfo{author}{{Rodi} J},
  \bibinfo{author}{{Rodriguez} J}, \bibinfo{author}{{S{\'a}nchez-Fern{\'a}ndez}
  C}, \bibinfo{author}{{Sunyaev} R} and  \bibinfo{author}{{Ubertini} P}
  (\bibinfo{year}{2020}).
\bibinfo{title}{{INTEGRAL Discovery of a Burst with Associated Radio Emission
  from the Magnetar SGR 1935+2154}}.
\bibinfo{journal}{{\em The Astrophysical Journal Letters}}
  \bibinfo{volume}{898} (\bibinfo{number}{2}), \bibinfo{eid}{L29}.

\bibtype{Article}%
\bibitem[{Mereghetti} et al.(2024)]{Mereghetti24}
\bibinfo{author}{{Mereghetti} S}, \bibinfo{author}{{Rigoselli} M},
  \bibinfo{author}{{Salvaterra} R}, \bibinfo{author}{{Pacholski} DP},
  \bibinfo{author}{{Rodi} JC}, \bibinfo{author}{{Gotz} D},
  \bibinfo{author}{{Arrigoni} E}, \bibinfo{author}{{D'Avanzo} P},
  \bibinfo{author}{{Adami} C}, \bibinfo{author}{{Bazzano} A},
  \bibinfo{author}{{Bozzo} E}, \bibinfo{author}{{Brivio} R},
  \bibinfo{author}{{Campana} S}, \bibinfo{author}{{Cappellaro} E},
  \bibinfo{author}{{Chenevez} J}, \bibinfo{author}{{De Luise} F},
  \bibinfo{author}{{Ducci} L}, \bibinfo{author}{{Esposito} P},
  \bibinfo{author}{{Ferrigno} C}, \bibinfo{author}{{Ferro} M},
  \bibinfo{author}{{Israel} GL}, \bibinfo{author}{{Le Floc'h} E},
  \bibinfo{author}{{Martin-Carrillo} A}, \bibinfo{author}{{Onori} F},
  \bibinfo{author}{{Rea} N}, \bibinfo{author}{{Reguitti} A},
  \bibinfo{author}{{Savchenko} V}, \bibinfo{author}{{Souami} D},
  \bibinfo{author}{{Tartaglia} L}, \bibinfo{author}{{Thuillot} W},
  \bibinfo{author}{{Tiengo} A}, \bibinfo{author}{{Tomasella} L},
  \bibinfo{author}{{Topinka} M}, \bibinfo{author}{{Turpin} D} and
  \bibinfo{author}{{Ubertini} P} (\bibinfo{year}{2024}).
\bibinfo{title}{{A magnetar giant flare in the nearby starburst galaxy M82}}.
\bibinfo{journal}{{\em Nature}} \bibinfo{volume}{629} (\bibinfo{number}{8010}):
  \bibinfo{pages}{58--61}.

\bibtype{Article}%
\bibitem[{Olausen} and {Kaspi}(2014)]{Olausen14}
\bibinfo{author}{{Olausen} SA} and  \bibinfo{author}{{Kaspi} VM}
  (\bibinfo{year}{2014}).
\bibinfo{title}{{The McGill Magnetar Catalog}}.
\bibinfo{journal}{{\em The Astrophysical Journal Supplement}}
  \bibinfo{volume}{212} (\bibinfo{number}{1}), \bibinfo{eid}{6}.

\bibtype{Article}%
\bibitem[{Pacini}(1967)]{pacini67}
\bibinfo{author}{{Pacini} F} (\bibinfo{year}{1967}).
\bibinfo{title}{{Energy Emission from a Neutron Star}}.
\bibinfo{journal}{{\em Nature}} \bibinfo{volume}{216} (\bibinfo{number}{5115}):
  \bibinfo{pages}{567--568}.

\bibtype{Article}%
\bibitem[{Pacini}(1968)]{pacini68}
\bibinfo{author}{{Pacini} F} (\bibinfo{year}{1968}).
\bibinfo{title}{{Rotating Neutron Stars, Pulsars and Supernova Remnants}}.
\bibinfo{journal}{{\em Nature}} \bibinfo{volume}{219} (\bibinfo{number}{5150}):
  \bibinfo{pages}{145--146}.

\bibtype{Article}%
\bibitem[{Pacini}(1971)]{pacini71}
\bibinfo{author}{{Pacini} F} (\bibinfo{year}{1971}).
\bibinfo{title}{{The Secular Decrease of Optical and X-Ray Luminosity of
  Pulsars}}.
\bibinfo{journal}{{\em The Astrophysical Journal}} \bibinfo{volume}{163}:
  \bibinfo{pages}{L17}.

\bibtype{Article}%
\bibitem[{Palmer} et al.(2005)]{palmer05}
\bibinfo{author}{{Palmer} DM}, \bibinfo{author}{{Barthelmy} S},
  \bibinfo{author}{{Gehrels} N}, \bibinfo{author}{{Kippen} RM},
  \bibinfo{author}{{Cayton} T}, \bibinfo{author}{{Kouveliotou} C},
  \bibinfo{author}{{Eichler} D}, \bibinfo{author}{{Wijers} RAMJ},
  \bibinfo{author}{{Woods} PM}, \bibinfo{author}{{Granot} J},
  \bibinfo{author}{{Lyubarsky} YE}, \bibinfo{author}{{Ramirez-Ruiz} E},
  \bibinfo{author}{{Barbier} L}, \bibinfo{author}{{Chester} M},
  \bibinfo{author}{{Cummings} J}, \bibinfo{author}{{Fenimore} EE},
  \bibinfo{author}{{Finger} MH}, \bibinfo{author}{{Gaensler} BM},
  \bibinfo{author}{{Hullinger} D}, \bibinfo{author}{{Krimm} H},
  \bibinfo{author}{{Markwardt} CB}, \bibinfo{author}{{Nousek} JA},
  \bibinfo{author}{{Parsons} A}, \bibinfo{author}{{Patel} S},
  \bibinfo{author}{{Sakamoto} T}, \bibinfo{author}{{Sato} G},
  \bibinfo{author}{{Suzuki} M} and  \bibinfo{author}{{Tueller} J}
  (\bibinfo{year}{2005}).
\bibinfo{title}{{A giant {$\gamma$}-ray flare from the magnetar
  SGR\,1806--20}}.
\bibinfo{journal}{{\em Nature}} \bibinfo{volume}{434}:
  \bibinfo{pages}{1107--1109}.

\bibtype{Article}%
\bibitem[{Pavlov} et al.(2000)]{Pavlov00}
\bibinfo{author}{{Pavlov} GG}, \bibinfo{author}{{Zavlin} VE},
  \bibinfo{author}{{Aschenbach} B}, \bibinfo{author}{{Tr{\"u}mper} J} and
  \bibinfo{author}{{Sanwal} D} (\bibinfo{year}{2000}).
\bibinfo{title}{{The Compact Central Object in Cassiopeia A: A Neutron Star
  with Hot Polar Caps or a Black Hole?}}
\bibinfo{journal}{{\em The Astrophysical Journal}} \bibinfo{volume}{531}
  (\bibinfo{number}{1}): \bibinfo{pages}{L53--L56}.

\bibtype{Article}%
\bibitem[{Petroff} et al.(2022)]{Petroff2022}
\bibinfo{author}{{Petroff} E}, \bibinfo{author}{{Hessels} JWT} and
  \bibinfo{author}{{Lorimer} DR} (\bibinfo{year}{2022}).
\bibinfo{title}{{Fast radio bursts at the dawn of the 2020s}}.
\bibinfo{journal}{{\em The Astronomy and Astrophysics Review}}
  \bibinfo{volume}{30} (\bibinfo{number}{1}), \bibinfo{eid}{2}.

\bibtype{Article}%
\bibitem[{Posselt} and {Pavlov}(2022)]{Posselt22}
\bibinfo{author}{{Posselt} B} and  \bibinfo{author}{{Pavlov} GG}
  (\bibinfo{year}{2022}).
\bibinfo{title}{{The Cooling of the Central Compact Object in Cas A from 2006
  to 2020}}.
\bibinfo{journal}{{\em The Astrophysical Journal}} \bibinfo{volume}{932}
  (\bibinfo{number}{2}), \bibinfo{eid}{83}.

\bibtype{Article}%
\bibitem[{Posselt} et al.(2018)]{Posselt18b}
\bibinfo{author}{{Posselt} B}, \bibinfo{author}{{Pavlov} GG},
  \bibinfo{author}{{Ertan} {\"U}}, \bibinfo{author}{{{\c{C}}al{\i}{\c{s}}kan}
  S}, \bibinfo{author}{{Luhman} KL} and  \bibinfo{author}{{Williams} CC}
  (\bibinfo{year}{2018}).
\bibinfo{title}{{Discovery of Extended Infrared Emission around the Neutron
  Star RXJ0806.4-4123}}.
\bibinfo{journal}{{\em The Astrophysical Journal}} \bibinfo{volume}{865}
  (\bibinfo{number}{1}), \bibinfo{eid}{1}.

\bibtype{Article}%
\bibitem[{Predehl} et al.(2021)]{predehl21}
\bibinfo{author}{{Predehl} P}, \bibinfo{author}{{Andritschke} R},
  \bibinfo{author}{{Arefiev} V}, \bibinfo{author}{{Babyshkin} V},
  \bibinfo{author}{{Batanov} O}, \bibinfo{author}{{Becker} W},
  \bibinfo{author}{{B{\"o}hringer} H}, \bibinfo{author}{{Bogomolov} A},
  \bibinfo{author}{{Boller} T}, \bibinfo{author}{{Borm} K},
  \bibinfo{author}{{Bornemann} W}, \bibinfo{author}{{Br{\"a}uninger} H},
  \bibinfo{author}{{Br{\"u}ggen} M}, \bibinfo{author}{{Brunner} H},
  \bibinfo{author}{{Brusa} M}, \bibinfo{author}{{Bulbul} E},
  \bibinfo{author}{{Buntov} M}, \bibinfo{author}{{Burwitz} V},
  \bibinfo{author}{{Burkert} W}, \bibinfo{author}{{Clerc} N},
  \bibinfo{author}{{Churazov} E}, \bibinfo{author}{{Coutinho} D},
  \bibinfo{author}{{Dauser} T}, \bibinfo{author}{{Dennerl} K},
  \bibinfo{author}{{Doroshenko} V}, \bibinfo{author}{{Eder} J},
  \bibinfo{author}{{Emberger} V}, \bibinfo{author}{{Eraerds} T},
  \bibinfo{author}{{Finoguenov} A}, \bibinfo{author}{{Freyberg} M},
  \bibinfo{author}{{Friedrich} P}, \bibinfo{author}{{Friedrich} S},
  \bibinfo{author}{{F{\"u}rmetz} M}, \bibinfo{author}{{Georgakakis} A},
  \bibinfo{author}{{Gilfanov} M}, \bibinfo{author}{{Granato} S},
  \bibinfo{author}{{Grossberger} C}, \bibinfo{author}{{Gueguen} A},
  \bibinfo{author}{{Gureev} P}, \bibinfo{author}{{Haberl} F},
  \bibinfo{author}{{H{\"a}lker} O}, \bibinfo{author}{{Hartner} G},
  \bibinfo{author}{{Hasinger} G}, \bibinfo{author}{{Huber} H},
  \bibinfo{author}{{Ji} L}, \bibinfo{author}{{Kienlin} Av},
  \bibinfo{author}{{Kink} W}, \bibinfo{author}{{Korotkov} F},
  \bibinfo{author}{{Kreykenbohm} I}, \bibinfo{author}{{Lamer} G},
  \bibinfo{author}{{Lomakin} I}, \bibinfo{author}{{Lapshov} I},
  \bibinfo{author}{{Liu} T}, \bibinfo{author}{{Maitra} C},
  \bibinfo{author}{{Meidinger} N}, \bibinfo{author}{{Menz} B},
  \bibinfo{author}{{Merloni} A}, \bibinfo{author}{{Mernik} T},
  \bibinfo{author}{{Mican} B}, \bibinfo{author}{{Mohr} J},
  \bibinfo{author}{{M{\"u}ller} S}, \bibinfo{author}{{Nandra} K},
  \bibinfo{author}{{Nazarov} V}, \bibinfo{author}{{Pacaud} F},
  \bibinfo{author}{{Pavlinsky} M}, \bibinfo{author}{{Perinati} E},
  \bibinfo{author}{{Pfeffermann} E}, \bibinfo{author}{{Pietschner} D},
  \bibinfo{author}{{Ramos-Ceja} ME}, \bibinfo{author}{{Rau} A},
  \bibinfo{author}{{Reiffers} J}, \bibinfo{author}{{Reiprich} TH},
  \bibinfo{author}{{Robrade} J}, \bibinfo{author}{{Salvato} M},
  \bibinfo{author}{{Sanders} J}, \bibinfo{author}{{Santangelo} A},
  \bibinfo{author}{{Sasaki} M}, \bibinfo{author}{{Scheuerle} H},
  \bibinfo{author}{{Schmid} C}, \bibinfo{author}{{Schmitt} J},
  \bibinfo{author}{{Schwope} A}, \bibinfo{author}{{Shirshakov} A},
  \bibinfo{author}{{Steinmetz} M}, \bibinfo{author}{{Stewart} I},
  \bibinfo{author}{{Str{\"u}der} L}, \bibinfo{author}{{Sunyaev} R},
  \bibinfo{author}{{Tenzer} C}, \bibinfo{author}{{Tiedemann} L},
  \bibinfo{author}{{Tr{\"u}mper} J}, \bibinfo{author}{{Voron} V},
  \bibinfo{author}{{Weber} P}, \bibinfo{author}{{Wilms} J} and
  \bibinfo{author}{{Yaroshenko} V} (\bibinfo{year}{2021}).
\bibinfo{title}{{The eROSITA X-ray telescope on SRG}}.
\bibinfo{journal}{{\em Astronomy and Astrophysics}} \bibinfo{volume}{647},
  \bibinfo{eid}{A1}.

\bibtype{Article}%
\bibitem[{Rea} et al.(2016)]{Rea16}
\bibinfo{author}{{Rea} N}, \bibinfo{author}{{Borghese} A},
  \bibinfo{author}{{Esposito} P}, \bibinfo{author}{{Coti Zelati} F},
  \bibinfo{author}{{Bachetti} M}, \bibinfo{author}{{Israel} GL} and
  \bibinfo{author}{{De Luca} A} (\bibinfo{year}{2016}).
\bibinfo{title}{{Magnetar-like Activity from the Central Compact Object in the
  SNR RCW103}}.
\bibinfo{journal}{{\em The Astrophysical Journal Letters}}
  \bibinfo{volume}{828} (\bibinfo{number}{1}), \bibinfo{eid}{L13}.

\bibtype{Article}%
\bibitem[{Rodr{\'\i}guez Castillo} et al.(2016)]{Rodriguez16}
\bibinfo{author}{{Rodr{\'\i}guez Castillo} GA}, \bibinfo{author}{{Israel} GL},
  \bibinfo{author}{{Tiengo} A}, \bibinfo{author}{{Salvetti} D},
  \bibinfo{author}{{Turolla} R}, \bibinfo{author}{{Zane} S},
  \bibinfo{author}{{Rea} N}, \bibinfo{author}{{Esposito} P},
  \bibinfo{author}{{Mereghetti} S}, \bibinfo{author}{{Perna} R},
  \bibinfo{author}{{Stella} L}, \bibinfo{author}{{Pons} JA},
  \bibinfo{author}{{Campana} S}, \bibinfo{author}{{G{\"o}tz} D} and
  \bibinfo{author}{{Motta} S} (\bibinfo{year}{2016}).
\bibinfo{title}{{The outburst decay of the low magnetic field magnetar SWIFT
  J1822.3-1606: phase-resolved analysis and evidence for a variable cyclotron
  feature}}.
\bibinfo{journal}{{\em Monthly Notices of the Royal Astronomical Society}}
  \bibinfo{volume}{456} (\bibinfo{number}{4}): \bibinfo{pages}{4145--4155}.

\bibtype{Inproceedings}%
\bibitem[{Shearer} and {Connor}(2018)]{Shearer2018}
\bibinfo{author}{{Shearer} A} and  \bibinfo{author}{{Connor} EO}
  (\bibinfo{year}{2018}), \bibinfo{title}{{Optical pulsars and polarimetry}},
  \bibinfo{editor}{{Weltevrede} P}, \bibinfo{editor}{{Perera} BBP},
  \bibinfo{editor}{{Preston} LL} and  \bibinfo{editor}{{Sanidas} S}, (Eds.),
  \bibinfo{booktitle}{Pulsar Astrophysics the Next Fifty Years},
  \bibinfo{volume}{337},  \bibinfo{pages}{191--194}.

\bibtype{Article}%
\bibitem[{Smith} et al.(2023)]{Smith2023}
\bibinfo{author}{{Smith} DA}, \bibinfo{author}{{Abdollahi} S},
  \bibinfo{author}{{Ajello} M}, \bibinfo{author}{{Bailes} M},
  \bibinfo{author}{{Baldini} L}, \bibinfo{author}{{Ballet} J},
  \bibinfo{author}{{Baring} MG}, \bibinfo{author}{{Bassa} C},
  \bibinfo{author}{{Gonzalez} JB}, \bibinfo{author}{{Bellazzini} R},
  \bibinfo{author}{{Berretta} A}, \bibinfo{author}{{Bhattacharyya} B},
  \bibinfo{author}{{Bissaldi} E}, \bibinfo{author}{{Bonino} R},
  \bibinfo{author}{{Bottacini} E}, \bibinfo{author}{{Bregeon} J},
  \bibinfo{author}{{Bruel} P}, \bibinfo{author}{{Burgay} M},
  \bibinfo{author}{{Burnett} TH}, \bibinfo{author}{{Cameron} RA},
  \bibinfo{author}{{Camilo} F}, \bibinfo{author}{{Caputo} R},
  \bibinfo{author}{{Caraveo} PA}, \bibinfo{author}{{Cavazzuti} E},
  \bibinfo{author}{{Chiaro} G}, \bibinfo{author}{{Ciprini} S},
  \bibinfo{author}{{Clark} CJ}, \bibinfo{author}{{Cognard} I},
  \bibinfo{author}{{Corongiu} A}, \bibinfo{author}{{Orestano} PC},
  \bibinfo{author}{{Crnogorcevic} M}, \bibinfo{author}{{Cuoco} A},
  \bibinfo{author}{{Cutini} S}, \bibinfo{author}{{D'Ammando} F},
  \bibinfo{author}{{de Angelis} A}, \bibinfo{author}{{DeCesar} ME},
  \bibinfo{author}{{De Gaetano} S}, \bibinfo{author}{{de Menezes} R},
  \bibinfo{author}{{Deneva} J}, \bibinfo{author}{{de Palma} F},
  \bibinfo{author}{{Di Lalla} N}, \bibinfo{author}{{Dirirsa} F},
  \bibinfo{author}{{Di Venere} L}, \bibinfo{author}{{Dom{\'\i}nguez} A},
  \bibinfo{author}{{Dumora} D}, \bibinfo{author}{{Fegan} SJ},
  \bibinfo{author}{{Ferrara} EC}, \bibinfo{author}{{Fiori} A},
  \bibinfo{author}{{Fleischhack} H}, \bibinfo{author}{{Flynn} C},
  \bibinfo{author}{{Franckowiak} A}, \bibinfo{author}{{Freire} PCC},
  \bibinfo{author}{{Fukazawa} Y}, \bibinfo{author}{{Fusco} P},
  \bibinfo{author}{{Galanti} G}, \bibinfo{author}{{Gammaldi} V},
  \bibinfo{author}{{Gargano} F}, \bibinfo{author}{{Gasparrini} D},
  \bibinfo{author}{{Giacchino} F}, \bibinfo{author}{{Giglietto} N},
  \bibinfo{author}{{Giordano} F}, \bibinfo{author}{{Giroletti} M},
  \bibinfo{author}{{Green} D}, \bibinfo{author}{{Grenier} IA},
  \bibinfo{author}{{Guillemot} L}, \bibinfo{author}{{Guiriec} S},
  \bibinfo{author}{{Gustafsson} M}, \bibinfo{author}{{Harding} AK},
  \bibinfo{author}{{Hays} E}, \bibinfo{author}{{Hewitt} JW},
  \bibinfo{author}{{Horan} D}, \bibinfo{author}{{Hou} X},
  \bibinfo{author}{{Jankowski} F}, \bibinfo{author}{{Johnson} RP},
  \bibinfo{author}{{Johnson} TJ}, \bibinfo{author}{{Johnston} S},
  \bibinfo{author}{{Kataoka} J}, \bibinfo{author}{{Keith} MJ},
  \bibinfo{author}{{Kerr} M}, \bibinfo{author}{{Kramer} M},
  \bibinfo{author}{{Kuss} M}, \bibinfo{author}{{Latronico} L},
  \bibinfo{author}{{Lee} SH}, \bibinfo{author}{{Li} D}, \bibinfo{author}{{Li}
  J}, \bibinfo{author}{{Limyansky} B}, \bibinfo{author}{{Longo} F},
  \bibinfo{author}{{Loparco} F}, \bibinfo{author}{{Lorusso} L},
  \bibinfo{author}{{Lovellette} MN}, \bibinfo{author}{{Lower} M},
  \bibinfo{author}{{Lubrano} P}, \bibinfo{author}{{Lyne} AG},
  \bibinfo{author}{{Maan} Y}, \bibinfo{author}{{Maldera} S},
  \bibinfo{author}{{Manchester} RN}, \bibinfo{author}{{Manfreda} A},
  \bibinfo{author}{{Marelli} M}, \bibinfo{author}{{Mart{\'\i}-Devesa} G},
  \bibinfo{author}{{Mazziotta} MN}, \bibinfo{author}{{McEnery} JE},
  \bibinfo{author}{{Mereu} I}, \bibinfo{author}{{Michelson} PF},
  \bibinfo{author}{{Mickaliger} M}, \bibinfo{author}{{Mitthumsiri} W},
  \bibinfo{author}{{Mizuno} T}, \bibinfo{author}{{Moiseev} AA},
  \bibinfo{author}{{Monzani} ME}, \bibinfo{author}{{Morselli} A},
  \bibinfo{author}{{Negro} M}, \bibinfo{author}{{Nemmen} R},
  \bibinfo{author}{{Nieder} L}, \bibinfo{author}{{Nuss} E},
  \bibinfo{author}{{Omodei} N}, \bibinfo{author}{{Orienti} M},
  \bibinfo{author}{{Orlando} E}, \bibinfo{author}{{Ormes} JF},
  \bibinfo{author}{{Palatiello} M}, \bibinfo{author}{{Paneque} D},
  \bibinfo{author}{{Panzarini} G}, \bibinfo{author}{{Parthasarathy} A},
  \bibinfo{author}{{Persic} M}, \bibinfo{author}{{Pesce-Rollins} M},
  \bibinfo{author}{{Pillera} R}, \bibinfo{author}{{Poon} H},
  \bibinfo{author}{{Porter} TA}, \bibinfo{author}{{Possenti} A},
  \bibinfo{author}{{Principe} G}, \bibinfo{author}{{Rain{\`o}} S},
  \bibinfo{author}{{Rando} R}, \bibinfo{author}{{Ransom} SM},
  \bibinfo{author}{{Ray} PS}, \bibinfo{author}{{Razzano} M},
  \bibinfo{author}{{Razzaque} S}, \bibinfo{author}{{Reimer} A},
  \bibinfo{author}{{Reimer} O}, \bibinfo{author}{{Renault-Tinacci} N},
  \bibinfo{author}{{Romani} RW}, \bibinfo{author}{{S{\'a}nchez-Conde} M},
  \bibinfo{author}{{Parkinson} PMS}, \bibinfo{author}{{Scotton} L},
  \bibinfo{author}{{Serini} D}, \bibinfo{author}{{Sgr{\`o}} C},
  \bibinfo{author}{{Shannon} R}, \bibinfo{author}{{Sharma} V},
  \bibinfo{author}{{Shen} Z}, \bibinfo{author}{{Siskind} EJ},
  \bibinfo{author}{{Spandre} G}, \bibinfo{author}{{Spinelli} P},
  \bibinfo{author}{{Stappers} BW}, \bibinfo{author}{{Stephens} TE},
  \bibinfo{author}{{Suson} DJ}, \bibinfo{author}{{Tabassum} S},
  \bibinfo{author}{{Tajima} H}, \bibinfo{author}{{Tak} D},
  \bibinfo{author}{{Theureau} G}, \bibinfo{author}{{Thompson} DJ},
  \bibinfo{author}{{Tibolla} O}, \bibinfo{author}{{Torres} DF},
  \bibinfo{author}{{Valverde} J}, \bibinfo{author}{{Venter} C},
  \bibinfo{author}{{Wadiasingh} Z}, \bibinfo{author}{{Wang} N},
  \bibinfo{author}{{Wang} N}, \bibinfo{author}{{Wang} P},
  \bibinfo{author}{{Weltevrede} P}, \bibinfo{author}{{Wood} K},
  \bibinfo{author}{{Yan} J}, \bibinfo{author}{{Zaharijas} G},
  \bibinfo{author}{{Zhang} C} and  \bibinfo{author}{{Zhu} W}
  (\bibinfo{year}{2023}).
\bibinfo{title}{{The Third Fermi Large Area Telescope Catalog of Gamma-Ray
  Pulsars}}.
\bibinfo{journal}{{\em The Astrophysical Journal}} \bibinfo{volume}{958}
  (\bibinfo{number}{2}), \bibinfo{eid}{191}.

\bibtype{Article}%
\bibitem[{Tan} et al.(2018)]{Tan2018}
\bibinfo{author}{{Tan} CM}, \bibinfo{author}{{Bassa} CG},
  \bibinfo{author}{{Cooper} S}, \bibinfo{author}{{Dijkema} TJ},
  \bibinfo{author}{{Esposito} P}, \bibinfo{author}{{Hessels} JWT},
  \bibinfo{author}{{Kondratiev} VI}, \bibinfo{author}{{Kramer} M},
  \bibinfo{author}{{Michilli} D}, \bibinfo{author}{{Sanidas} S},
  \bibinfo{author}{{Shimwell} TW}, \bibinfo{author}{{Stappers} BW},
  \bibinfo{author}{{van Leeuwen} J}, \bibinfo{author}{{Cognard} I},
  \bibinfo{author}{{Grie{\ss}meier} JM}, \bibinfo{author}{{Karastergiou} A},
  \bibinfo{author}{{Keane} EF}, \bibinfo{author}{{Sobey} C} and
  \bibinfo{author}{{Weltevrede} P} (\bibinfo{year}{2018}).
\bibinfo{title}{{LOFAR Discovery of a 23.5 s Radio Pulsar}}.
\bibinfo{journal}{{\em The Astrophysical Journal}} \bibinfo{volume}{866}
  (\bibinfo{number}{1}), \bibinfo{eid}{54}.

\bibtype{Article}%
\bibitem[{Taverna} and {Turolla}(2024)]{Taverna2024}
\bibinfo{author}{{Taverna} R} and  \bibinfo{author}{{Turolla} R}
  (\bibinfo{year}{2024}).
\bibinfo{title}{{X-ray Polarization from Magnetar Sources}}.
\bibinfo{journal}{{\em Galaxies}} \bibinfo{volume}{12} (\bibinfo{number}{1}),
  \bibinfo{eid}{6}.

\bibtype{Article}%
\bibitem[{Tiengo} et al.(2013)]{Tiengo13}
\bibinfo{author}{{Tiengo} A}, \bibinfo{author}{{Esposito} P},
  \bibinfo{author}{{Mereghetti} S}, \bibinfo{author}{{Turolla} R},
  \bibinfo{author}{{Nobili} L}, \bibinfo{author}{{Gastaldello} F},
  \bibinfo{author}{{G{\"o}tz} D}, \bibinfo{author}{{Israel} GL},
  \bibinfo{author}{{Rea} N}, \bibinfo{author}{{Stella} L},
  \bibinfo{author}{{Zane} S} and  \bibinfo{author}{{Bignami} GF}
  (\bibinfo{year}{2013}).
\bibinfo{title}{{A variable absorption feature in the X-ray spectrum of a
  magnetar}}.
\bibinfo{journal}{{\em Nature}} \bibinfo{volume}{500} (\bibinfo{number}{7462}):
  \bibinfo{pages}{312--314}.

\bibtype{Article}%
\bibitem[{Torne} et al.(2022)]{torne22}
\bibinfo{author}{{Torne} P}, \bibinfo{author}{{Bell} GS},
  \bibinfo{author}{{Bintley} D}, \bibinfo{author}{{Desvignes} G},
  \bibinfo{author}{{Berry} D}, \bibinfo{author}{{Dempsey} JT},
  \bibinfo{author}{{Ho} PTP}, \bibinfo{author}{{Parsons} H},
  \bibinfo{author}{{Eatough} RP}, \bibinfo{author}{{Karuppusamy} R},
  \bibinfo{author}{{Kramer} M}, \bibinfo{author}{{Kramer} C},
  \bibinfo{author}{{Liu} K}, \bibinfo{author}{{Paubert} G},
  \bibinfo{author}{{Sanchez-Portal} M} and  \bibinfo{author}{{Schuster} KF}
  (\bibinfo{year}{2022}).
\bibinfo{title}{{Submillimeter Pulsations from the Magnetar XTE J1810-197}}.
\bibinfo{journal}{{\em The Astrophysical Journal Letters}}
  \bibinfo{volume}{925} (\bibinfo{number}{2}), \bibinfo{eid}{L17}.

\bibtype{Inproceedings}%
\bibitem[{Turolla}(2009)]{Turolla09}
\bibinfo{author}{{Turolla} R} (\bibinfo{year}{2009}), \bibinfo{title}{{Isolated
  Neutron Stars: The Challenge of Simplicity}}, \bibinfo{editor}{{Becker} W},
  (Ed.), \bibinfo{booktitle}{Astrophysics and Space Science Library},
  \bibinfo{series}{Astrophysics and Space Science Library},
  \bibinfo{volume}{357}, pp. \bibinfo{pages}{141}.

\bibtype{Article}%
\bibitem[{Turolla} and {Esposito}(2013)]{Turolla13}
\bibinfo{author}{{Turolla} R} and  \bibinfo{author}{{Esposito} P}
  (\bibinfo{year}{2013}).
\bibinfo{title}{{Low-Magnetic Magnetars}}.
\bibinfo{journal}{{\em International Journal of Modern Physics D}}
  \bibinfo{volume}{22} (\bibinfo{number}{13}), \bibinfo{eid}{1330024-163}.

\bibtype{Article}%
\bibitem[{van Paradijs} et al.(1995)]{vanparadijs95}
\bibinfo{author}{{van Paradijs} J}, \bibinfo{author}{{Taam} RE} and
  \bibinfo{author}{{van den Heuvel} EPJ} (\bibinfo{year}{1995}),
  \bibinfo{month}{Jul.}
\bibinfo{title}{{On the nature of the 'anomalous' 6-s X-ray pulsars}}.
\bibinfo{journal}{{\em Astronomy and Astrophysics}} \bibinfo{volume}{299}:
  \bibinfo{pages}{L41}.

\bibtype{Article}%
\bibitem[{Wadiasingh} et al.(2018)]{Wadiasingh18}
\bibinfo{author}{{Wadiasingh} Z}, \bibinfo{author}{{Baring} MG},
  \bibinfo{author}{{Gonthier} PL} and  \bibinfo{author}{{Harding} AK}
  (\bibinfo{year}{2018}).
\bibinfo{title}{{Resonant Inverse Compton Scattering Spectra from Highly
  Magnetized Neutron Stars}}.
\bibinfo{journal}{{\em The Astrophysical Journal}} \bibinfo{volume}{854}
  (\bibinfo{number}{2}), \bibinfo{eid}{98}.

\bibtype{Article}%
\bibitem[{Wang} et al.(2020)]{wang20}
\bibinfo{author}{{Wang} HH}, \bibinfo{author}{{Lin} LCC},
  \bibinfo{author}{{Dai} S}, \bibinfo{author}{{Takata} J},
  \bibinfo{author}{{Li} KL}, \bibinfo{author}{{Hu} CP} and
  \bibinfo{author}{{Hou} X} (\bibinfo{year}{2020}).
\bibinfo{title}{{A Multiwavelength Study of PSR J1119-6127 after 2016
  Outburst}}.
\bibinfo{journal}{{\em The Astrophysical Journal}} \bibinfo{volume}{902}
  (\bibinfo{number}{2}), \bibinfo{eid}{96}.

\bibtype{Article}%
\bibitem[{Wex} and {Kramer}(2020)]{Wex2020}
\bibinfo{author}{{Wex} N} and  \bibinfo{author}{{Kramer} M}
  (\bibinfo{year}{2020}).
\bibinfo{title}{{Gravity Tests with Radio Pulsars}}.
\bibinfo{journal}{{\em Universe}} \bibinfo{volume}{6} (\bibinfo{number}{9}),
  \bibinfo{eid}{156}.

\bibtype{Article}%
\bibitem[{Younes} et al.(2020{\natexlab{a}})]{Younes20}
\bibinfo{author}{{Younes} G}, \bibinfo{author}{{G{\"u}ver} T},
  \bibinfo{author}{{Kouveliotou} C}, \bibinfo{author}{{Baring} MG},
  \bibinfo{author}{{Hu} CP}, \bibinfo{author}{{Wadiasingh} Z},
  \bibinfo{author}{{Begi{\c{c}}arslan} B}, \bibinfo{author}{{Enoto} T},
  \bibinfo{author}{{G{\"o}{\u{g}}{\"u}{\c{s}}} E}, \bibinfo{author}{{Lin} L},
  \bibinfo{author}{{Harding} AK}, \bibinfo{author}{{van der Horst} AJ},
  \bibinfo{author}{{Majid} WA}, \bibinfo{author}{{Guillot} S} and
  \bibinfo{author}{{Malacaria} C} (\bibinfo{year}{2020}{\natexlab{a}}).
\bibinfo{title}{{NICER View of the 2020 Burst Storm and Persistent Emission of
  SGR 1935+2154}}.
\bibinfo{journal}{{\em The Astrophysical Journal Letters}}
  \bibinfo{volume}{904} (\bibinfo{number}{2}), \bibinfo{eid}{L21}.

\bibtype{Article}%
\bibitem[{Younes} et al.(2020{\natexlab{b}})]{Younes2020}
\bibinfo{author}{{Younes} G}, \bibinfo{author}{{Ray} PS},
  \bibinfo{author}{{Baring} MG}, \bibinfo{author}{{Kouveliotou} C},
  \bibinfo{author}{{Fletcher} C}, \bibinfo{author}{{Wadiasingh} Z},
  \bibinfo{author}{{Harding} AK} and  \bibinfo{author}{{Goldstein} A}
  (\bibinfo{year}{2020}{\natexlab{b}}).
\bibinfo{title}{{A Radiatively Quiet Glitch and Anti-glitch in the Magnetar 1E
  2259+586}}.
\bibinfo{journal}{{\em The Astrophysical Journal Letters}}
  \bibinfo{volume}{896} (\bibinfo{number}{2}), \bibinfo{eid}{L42}.

\bibtype{Article}%
\bibitem[{Younes} et al.(2023)]{Younes2023}
\bibinfo{author}{{Younes} G}, \bibinfo{author}{{Baring} MG},
  \bibinfo{author}{{Harding} AK}, \bibinfo{author}{{Enoto} T},
  \bibinfo{author}{{Wadiasingh} Z}, \bibinfo{author}{{Pearlman} AB},
  \bibinfo{author}{{Ho} WCG}, \bibinfo{author}{{Guillot} S},
  \bibinfo{author}{{Arzoumanian} Z}, \bibinfo{author}{{Borghese} A},
  \bibinfo{author}{{Gendreau} K}, \bibinfo{author}{{G{\"o}{\v{g}}{\"u}{\c{s}}}
  E}, \bibinfo{author}{{G{\"u}ver} T}, \bibinfo{author}{{van der Horst} AJ},
  \bibinfo{author}{{Hu} CP}, \bibinfo{author}{{Jaisawal} GK},
  \bibinfo{author}{{Kouveliotou} C}, \bibinfo{author}{{Lin} L} and
  \bibinfo{author}{{Majid} WA} (\bibinfo{year}{2023}).
\bibinfo{title}{{Magnetar spin-down glitch clearing the way for FRB-like bursts
  and a pulsed radio episode}}.
\bibinfo{journal}{{\em Nature Astronomy}} \bibinfo{volume}{7}:
  \bibinfo{pages}{339--350}.

\bibtype{Article}%
\bibitem[{Zhu} et al.(2023)]{Zhu23}
\bibinfo{author}{{Zhu} W}, \bibinfo{author}{{Xu} H}, \bibinfo{author}{{Zhou}
  D}, \bibinfo{author}{{Lin} L}, \bibinfo{author}{{Wang} B},
  \bibinfo{author}{{Wang} P}, \bibinfo{author}{{Zhang} C},
  \bibinfo{author}{{Niu} J}, \bibinfo{author}{{Chen} Y}, \bibinfo{author}{{Li}
  C}, \bibinfo{author}{{Meng} L}, \bibinfo{author}{{Lee} K},
  \bibinfo{author}{{Zhang} B}, \bibinfo{author}{{Feng} Y},
  \bibinfo{author}{{Ge} M}, \bibinfo{author}{{G{\"o}{\u{g}}{\"u}{\c{s}}} E},
  \bibinfo{author}{{Guan} X}, \bibinfo{author}{{Han} J},
  \bibinfo{author}{{Jiang} J}, \bibinfo{author}{{Jiang} P},
  \bibinfo{author}{{Kouveliotou} C}, \bibinfo{author}{{Li} D},
  \bibinfo{author}{{Miao} C}, \bibinfo{author}{{Miao} X},
  \bibinfo{author}{{Men} Y}, \bibinfo{author}{{Niu} C}, \bibinfo{author}{{Wang}
  W}, \bibinfo{author}{{Wang} Z}, \bibinfo{author}{{Xu} J},
  \bibinfo{author}{{Xu} R}, \bibinfo{author}{{Xue} M}, \bibinfo{author}{{Yang}
  Y}, \bibinfo{author}{{Yu} W}, \bibinfo{author}{{Yuan} M},
  \bibinfo{author}{{Yue} Y}, \bibinfo{author}{{Zhang} S} and
  \bibinfo{author}{{Zhang} Y} (\bibinfo{year}{2023}).
\bibinfo{title}{{A radio pulsar phase from SGR J1935+2154 provides clues to the
  magnetar FRB mechanism}}.
\bibinfo{journal}{{\em Science Advances}} \bibinfo{volume}{9}
  (\bibinfo{number}{30}), \bibinfo{eid}{eadf6198}.

\end{thebibliography*}

\end{document}